\documentclass[12pt,preprint]{aastex}
\usepackage{amssymb}
\usepackage{amsmath}
\usepackage{epstopdf}
\usepackage{natbib}
\usepackage[colorlinks, citecolor=blue]{hyperref}
\usepackage{hyperref}
\usepackage[all]{hypcap}

\begin{document}

\title{A Unified Energy-Reservoir Model Containing Contributions from $^{56}$Ni and Neutron Stars and Its Implication to Luminous Type Ic Supernovae}
\author{S. Q. Wang\altaffilmark{1,2}, L. J. Wang\altaffilmark{1,2}, Z. G. Dai\altaffilmark{1,2} and X. F. Wu\altaffilmark{3,4,5}}
\affil{\altaffilmark{1}School of Astronomy and Space Science, Nanjing University, Nanjing 210093, China; dzg@nju.edu.cn}
\affil{\altaffilmark{2}Key Laboratory of Modern Astronomy and Astrophysics (Nanjing University), Ministry of Education, China}
\affil{\altaffilmark{3}Purple Mountain Observatory, Chinese Academy of Sciences, Nanjing, 210008, China}
\affil{\altaffilmark{4}Chinese Center for Antarctic Astronomy, Chinese Academy of Sciences, Nanjing, 210008, China}
\affil{\altaffilmark{5}Joint Center for Particle Nuclear Physics and Cosmology of Purple Mountain
Observatory-Nanjing University, Chinese Academy of Sciences, Nanjing 210008, China}

\begin{abstract}

Most type-Ic core-collapse supernovae (CCSNe) produce $^{56}$Ni and neutron stars (NSs) or black holes (BHs).
 The dipole radiation of nascent NSs has usually been neglected in explaining
 supernovae (SNe) with peak absolute magnitude $M_{\rm peak}$ in any band are $\gtrsim -19.5$~mag, while the
 $^{56}$Ni can be neglected in fitting
 most type-Ic superluminous supernovae (SLSNe Ic) whose $M_{\rm peak}$ in any band are $\lesssim -21$~mag,
 since the luminosity from a magnetar (highly magnetized NS)
  can outshine that from a moderate amount of $^{56}$Ni.
For luminous SNe Ic with $-21 \lesssim M_{\rm peak}\lesssim -19.5$~mag, however,
 both contributions from $^{56}$Ni and NSs cannot be neglected without serious modeling, since they are not SLSNe and
 the $^{56}$Ni mass could be up to $\sim 0.5 M_{\odot}$.
In this paper we propose a unified model that contain contributions from both $^{56}$Ni and a nascent NS.
We select three luminous SNe Ic-BL,
SN~2010ay, SN~2006nx, and SN~14475, and
show that, if these SNe are powered by $^{56}$Ni,
 the ratio of $M_{\rm Ni}$ to $M_{\rm ej}$ are unrealistic.
Alternatively, we invoke the magnetar model and the hybrid ($^{56}$Ni + NS) model
 and find that they can fit the observations,
 indicating that our models are valid and necessary for luminous SNe Ic.
Owing to the lack of late-time photometric data,
we cannot break the parameter degeneracy and thus distinguish among the model parameters,
 but we can expect that future multi-epoch 
 observations of luminous SNe can provide stringent constraints on $^{56}$Ni yields
 and the parameters of putative magnetars.

\end{abstract}


\section{Introduction}

It has long been believed that most aged massive (zero-age main-sequence mass $M_{\rm ZAMS}\gtrsim 8.0M_{\odot}$)
 stars terminate their lives as core-collapse supernovae (CCSNe) \citep{Woo2002,Jan2007},
which are classified into type IIP, IIL, IIn, IIb, Ib and Ic
according to their spectra and light curves \citep{Fil1997}, leaving neutron stars (NSs) or black holes (BHs) at the center and
 producing a moderate amount of $^{56}$Ni which is generally regarded as the dominant power source for most supernovae (SNe)
 \citep{Col1969,Col1980,Arn1982} \footnote{Since the observations for SN~1987A
  \citep{Lun2001} and SN~1998bw \citep{Sol2002} have already revealed
 that the radioactive elements other than $^{56}$Ni
 (e.g., $^{57}$Ni, $^{44}$Ti and $^{22}$Na, etc) constitute only a minor fraction of the radioactive masses
 and contribute dominant fluxes at very late times ($\gtrsim $600~days and 1,200$-$1,400~day after explosion for
 SN~1987A and SN~1998bw, respectively), we only consider the contribution of $^{56}$Ni, which is the most abundant nucleus resulting
  from explosive silicon burning in shock-heated silicon shells and the dominant energy source at early-time (e.g., $\leq500$ days) of a SN.}.
 Of all these subclasses, SNe Ic have attracted more and more attentions
since a great number of these events  have been discovered and confirmed in recent years that some SNe Ic
 with broad absorption line features (``broad-lined" or ``BL") have an accompanying gamma-ray burst (GRB) or X-ray flash (XRF) \citep{Woo2006,Hjo2012}.
 Main models of central engines of GRBs associated with SNe are the ``collapsar" model \citep{Woos1993,Mac1999} involving ``BH + disk" systems
 and the magnetar (highly magnetized neutron star) model \citep{Uso1992,Met2007,Buc2008,Met2011} proposing that the explosions may
 leave benind fast-rotating magnetars, being also regarded as the
 origin of the shallow decays and plateaus as well as rebrightenings in the multi-band afterglows of some GRBs
\citep{Dai1998a,Dai1998b,Zhang2001,Dai2004,Dai2012}.

On the other hand, in the last two decades, optical$-$NIR and radio observations have revealed that not every SN Ic-BL associates with
 a GRB or an XRF \citep{Sod2005,Dro2011}. Many SNe Ic-BL, no matter whether they are associated a(n) GRB/XRF or not, have very
 high kinetic energy $E_{\rm K}\gtrsim 1.0 \times10^{52}$~erg and have therefore been called
  ``Hypernovae" \citep{Iwa1998}. Supposing that the optical$-$NIR emission is powered by radioactive $^{56}$Ni decay,
 these SNe Ic need $\sim 0.1-0.5 M_\odot$ of $^{56}$Ni.
 All GRBs associated SNe Ic and most GRB-less SNe Ic are not very luminous, with peak absolute magnitude $M_{\rm peak}\gtrsim -19.5$.

Thanks to the unprecedented boom of targeted- and untargeted- sky survey programs, many
superluminous SNe (SLSNe) whose $M_{\rm peak}$ in any band are $\lesssim -21$~mag \citep{Gal2012} have been found
 in the past decade, most of which
 cannot be explained by the widely adopted $^{56}$Ni decay model \citep{Qui2011,Inse2013,Nich2013,McC2014,Nich2014},
 motivating researchers to consider alternative
 energy-reservoir models.
 Currently, main models explaining the SLSNe
 are SN ejecta - circumstellar medium (CSM) interaction model \citep{Che1982,Che1994,Che2011,Gin2012} that
 has been employed to explain many Type IIn luminous SNe \citep{Chu1994,Zhang2012,Ofe2013}
 and superluminous SNe IIn \citep{Smi2007,Mor2013} as well as some Type Ic SLSNe \citep{Cha2013,Nich2014},
 and the magnetar-powered SLSNe model \citep{Kas2010,Woos2010} that has been
 used to explain many Type Ic SLSNe \citep{Inse2013,Nich2013,How2013,McC2014,Vre2014,Nich2014,Wang2015}.
In some cases, the magnetar-powered SLSNe model with the assumption of full energy trapping
 fails to fit the late-time light curves \citep{Inse2013,Nich2014}.
 In \citet{Wang2015}, we generalized the magnetar-powered SLSNe model by introducing the hard emission leakage and
 solved this problem, highlighting the importance of the leakage effect in this model.

Thus, the power sources of most normal SNe and SLSNe have been
 attributed to $^{56}$Ni-decay and the magnetar spin-down or ejecta-CSM interaction, respectively. Within this picture,
 the dipole radiation of nascent magnetars has usually been neglected in explaining
 the SNe with $M_{\rm peak}\gtrsim -19.5$ \citep[but see][]{Mae2007}, while the
 $^{56}$Ni decay energy have generally been ignored in fitting
 all Type Ic SLSNe ($M_{\rm peak} \lesssim -21$)
 because the peak-luminosity due to the energy injection from a newly born fast-rotating (initial period $\sim$ millisecond) magnetar
  can outshine that from a moderate amount of $^{56}$Ni \citep[see][for further analysis]{Inse2013}.

Besides normal SNe and SLSNe mentioned above, however, there are some SNe
 with $-21\lesssim M_{\rm peak}\lesssim-19.5$~mag,
 constituting a class of ``gap-filler" events that bridge normal SNe and SLSNe,
 have been found by many telescopes. Most of these ``gap-filler" SNe are explained by the
 $^{56}$Ni decay model for SNe Ic
 as well as ``Super-Chandrasekhar-Mass" SNe Ia, 
 and the ejecta-CSM interaction model for SNe IIn.

Among these ``gap-filler" SNe with $-21\lesssim M_{\rm peak}\lesssim-19.5$~mag,
 luminous Type Ic SNe and their energy-reservoir mechanisms have not attracted enough attentions yet,
 their high peak-luminosities are simply attributed to the $^{56}$Ni cascade decay without any detailed modeling.
 It has recently been demonstrated that the $^{56}$Ni-decay model
 cannot be arbitrarily used to explain all SNe Ic especially those very luminous ones \citep{Inse2013,Nich2013} since
 the production of $^{56}$Ni is ineffective in CCSNe and
more $^{56}$Ni mass needs more mass of the ejecta, which could theoretically result in broad light curves
 that usually conflict with observations. While the CSM-interaction model can predict a wide range of peak luminosities and
 therefore explain the normal, luminous, and superluminous SNe IIn as a whole category,
 $^{56}$Ni-decay model cannot explain luminous SNe Ic.

Here, we propose that for some luminous SNe Ic, there are other energy reservoirs that play a significant
 role in producing light curves. Since all hydrogen envelopes of progenitors of SNe Ic have been stripped and the spectra are lack of narrow and/or
 emission lines indicative of the interactions \footnote{the unique exception is SN~2010mb
 which cannot be explained by $^{56}$Ni-powered model and has a signature of interactions \citep{Ben-A2014}}, we can exclude the
ionized hydrogen re-combination \citep{Fal1977,Kle1978,Pop1993,Kas2009} and do not consider the ejecta-CSM interaction process.
Therefore, like the cases of SLSNe Ic, the nascent magnetar embedded in the center of the explosion is the most promising candidate for
the additional power source \citep{Ost1971}.
On the other hand, the production of $^{56}$Ni is inevitable and cannot be
 neglected in modeling luminous SNe Ic because their peak luminosities are considerably
 lower than that of SLSNe and the contributions from a moderate amount of $^{56}$Ni can be comparable with
 the contributions given by any other energy reservoirs.
Furthermore, in principle, the contributions from $^{56}$Ni and NSs cannot be directly neglected for
 CCSNe with a wide range of peak luminosities. Thus, it is necessary to consider a unified model
 containing contributions from both $^{56}$Ni and NSs.

In this paper we construct the unified model which contain the contributions from both the $^{56}$Ni cascade decay energy and
 the NS rotational energy, and apply it to explain the light curves of some luminous SNe Ic-BL, i.e., SN~2010ay, SN~2006nx, and SN~14475.
 The remainder of this paper is organized as follows. In Section \ref{sec:uni}, we give
 a unified semi-analytical model for Type Ic SNe. Based on this model,
we fit the light curves of SN~2010ay, SN~2006nx, and SN~14475 in Section \ref{sec:fit}.
Finally, some discussions and conclusions are presented in Section \ref{sec:con}.

\section{The unified semi-analytical model for SNe}
\label{sec:uni}

In this section, we construct a unified semi-analytical model combining energy from a spin-down magnetar and an amount of $^{56}$Ni to
 describe the SN luminosity evolution.
In this unified model, $^{56}$Ni release high energy photons via the cascade decay chain $^{56}$Ni$\rightarrow$$^{56}$Co$\rightarrow$$^{56}$Fe,
 while the magnetar can inject its rotational energy into the SN ejecta as heat energy \citep{Ost1971}.

Based on \citet{Arn1982}, taking into account the $\gamma$-ray and X-ray leakage \citep[e.g.,][]{Clo1997,Val2008,Cha2009,Cha2012,Dro2013,Wang2015},
the luminosity is given by
\begin{eqnarray}
L(t)&=&\frac{2}{\tau_{m}}e^{-\left(\frac{t^{2}}{\tau_{m}^{2}}+\frac{2R_{0}t}{v\tau_{m}^{2}}\right)}~
\left(1-e^{-\tau_{\gamma}(t)}\right)\int_0^t e^{\left(\frac{t'^{2}}{\tau_{m}^{2}}+\frac{2R_{0}t'}{v\tau_{m}^{2}}\right)}   \nonumber\\
     &&\times\left(\frac{R_{0}}{v\tau_{m}}+\frac{t'}{\tau_{m}}\right)P(t')dt'~\mbox{erg s}^{-1},
\label{equ:lum}
\end{eqnarray}
where $R_{0}$ is the initial radius of the progenitor, which is very small compared to the radius of the ejecta. We take the limit $R_{0}\rightarrow0$,
 then the above equation can be largely simplified.
With Equations (18), (19) and (22) of \citet{Arn1982}, the effective light-curve timescale $\tau_{m}$ can be written as
 \footnote{The coefficient $2$
 in the equation has been adopted as $\frac{10}{3}$ in some other papers \citep{Cha2009,Cha2012,Cha2013,Wang2015},
 the latter originated from a typo in Equation (54) of \citet{Arn1982} that written $5/3$ as
 $3/5$, see also the footnote 1 in \citet{Whe2014}. To get the same light curves, $\kappa$ or $M_{\rm ej}$ ($v$)
should be multiplied by 5/3 (3/5), or these three parameters simultaneously adjusted
 so that $\tau_{m}$ can be invariant for an individual event.}
\begin{eqnarray}
\tau_{m}=\left(\frac{2\kappa M_{\rm ej}}{\beta vc}\right)^{1/2},
\label{equ:tau_m}
\end{eqnarray}
where $\kappa$ is the optical opacity to optical photons (i.e., the Thomson electron scattering opacity),
$M_{\rm ej}$ and $v$ are the mass and expansion speed of the ejecta, respectively, and $c$ is the speed of light.
$\beta \simeq 13.8$ is a constant that accounts for the density distribution of the ejecta. $P(t)$ is the power function. Here, $v$ is the scale velocity ($v_{\rm sc}$) in \citet{Arn1982}
 and approximates to the photospheric expansion velocity $v_{\rm ph}$. Hereafter, we let $v \simeq v_{\rm ph}$.

The factors $e^{-\tau_{\gamma}(t)}$ and $(1-e^{-\tau_{\gamma}(t)})$ in Equation (\ref{equ:lum}) represent the $\gamma$-ray leakage and trapping rate, respectively.
 $\tau_{\gamma}(t)~=~At^{-2}$ is the optical depth to $\gamma$-rays \citep{Cha2009,Cha2012}.
 If the SN ejecta has a uniform density distribution
 ($M_{\rm ej}~=~(4/3) \pi \rho R^3$, $E_{\rm K}~=~(3/10)M_{\rm ej}v^2$),
  $A$ depends on $\kappa_{\gamma}$ (the opacity to $\gamma$-rays), $M_{\rm ej}$ and $v$ as
\begin{eqnarray}
A&=&\frac{3\kappa_{\gamma} M_{\rm ej}}{4\pi v^2 }\nonumber \\
 &=&4.75 \times 10^{13}\left(\frac{\kappa_{\gamma}}{0.1~{\rm cm}^2~{\rm g}^{-1}}\right)  \nonumber \\
 &&\times\left(\frac{M_{\rm ej}}{M_\odot}\right)\left(\frac{v}{10^9~{\rm cm}~{\rm s}^{-1}}\right)^{-2}~\mbox{s}^{2},
\label{equ:leakage}
\end{eqnarray}

\subsection{The $^{56}$Ni-decay energy model}

In the $^{56}$Ni-decay energy model, the input power is
\begin{eqnarray}
&&P(t)=P_{\rm Ni}(t)= \epsilon_{\rm Ni}M_{\rm Ni}e^{-{t}/{\tau_{\rm Ni}}} + \epsilon_{\rm Co}M_{\rm Ni}\frac{e^{-{t}/{\tau_{\rm Co}}}-e^{-{t}/{\tau_{\rm Ni}}}}{1-{\tau_{\rm Ni}}/{\tau_{\rm Co}}}~\mbox{erg s}^{-1},
\label{equ:input-Ni}
\end{eqnarray}
where $\epsilon_{\rm Ni}~=~3.9\times 10^{10}$~erg~s$^{-1}$~g$^{-1}$ is the energy generation rate per unit mass due to $^{56}$Ni decay
($^{56}$Ni$\rightarrow$$^{56}$Co) \citep{Cap1997,Sut1984}, $M_{\rm Ni}$ is the initial mass of $^{56}$Ni,
 $\tau_{\rm Ni}~=~$8.8~days is the $e$-folding time of the $^{56}$Ni decay,
 $\epsilon_{\rm Co}=6.8 \times 10^{9}$ erg~s$^{-1}$~g$^{-1}$ is the energy generation rate due to $^{56}$Co decay
($^{56}$Co $\rightarrow$ $^{56}$Fe) \citep{Mae2003}, $\tau_{\rm Co}~=~$111.3~days
is the $e$-folding time of the $^{56}$Co decay.

\subsection{The magnetar spin-down energy model}

In the magnetar model, supposing that all the spin-down energy released
can be converted into the heat energy of SN ejecta and the angle between the dipole magnetic fields and the magnetar's spin axis is 45$^\circ$,
 the power coming from the magnetar dipole radiation is \citep{Ost1971}
\begin{equation}
P(t)=P_{\rm NS}(t) = \frac{E_{\rm NS}}{\tau_{\rm NS}} \frac{1}{(1+t/\tau_{\rm NS})^{2}}~\mbox{erg s}^{-1},
\label{equ:input-mag}
\end{equation}
$\tau_{\rm NS} = {6I_{\rm NS} c^3}/{B^2R_{\rm NS}^6\Omega_0^2} = 1.3~(B/10^{14}~{\rm G})^{-2}(P_0/10~{\rm ms})^2~ {\rm yr}$ is the spin-down timescale of the magnetar \citep{Kas2010}.
$E_{\rm NS}$ is the rotational energy of the magnetar,
\begin{eqnarray}
E_{\rm NS}&\simeq&({1}/{2})I_{\rm NS}\Omega_{\rm NS}^2 \nonumber \\
&\simeq&2\times 10^{52}\frac{M_{\rm NS}}{1.4~M_\odot}\left(\frac{P_{0}}{1~{\rm ms}}\right)^{-2}\left(\frac{R_{\rm NS}}{10~{\rm km}}\right)^2~\mbox{erg},
\end{eqnarray}
where $M_{\rm NS}$, $R_{\rm NS}$ and $P_{0}$ are the mass, radius and initial rotational period of the magnetar, respectively, while
$I_{\rm NS}=(2/5)M_{\rm NS}R_{\rm NS}^{2}$ is the moment of inertia of the magnetar
whose canonical value is $\sim 10^{45}$~g~cm$^{2}$ \citep{Woos2010}.

\subsection{The hybrid ($^{56}$Ni + magnetar) model}

If an amount of $^{56}$Ni is synthesized and a fast-rotating magnetar is left after a SN explosion,
 both contributions from these two power sources must be taken into account, this is the hybrid model.
In this model, the luminosity of a SN is
\begin{equation}
L_{\rm uni}(t)=L_{\rm Ni}(t)+L_{\rm NS}(t),
\label{equ:L-hyb}
\end{equation}
where $L_{\rm Ni}(t)$ and $L_{\rm NS}(t)$ are luminosities supplied by $^{56}$Ni and the magnetar, respectively.

In this paper we propose the ``unified'' model that contains the above three models, i.e., the $^{56}$Ni model, the magnetar model, and the hybrid model. If the contribution from the magnetar or $^{56}$Ni
 can be neglected, this model would be simplified to the magnetar model or the $^{56}$Ni-decay model, respectively.
The parameter set for the unified model is ($M_{\rm ej}$, $v$, $M_{\rm Ni}$, $B$, $P_0$, $\kappa_{\gamma}$). When
$M_{\rm Ni}=0$, the parameter set is ($M_{\rm ej}$, $v$, $B$, $P_0$, $\kappa_{\gamma}$), corresponding to the magnetar model;
when the neutron star is non-rotational, the parameter set is ($M_{\rm ej}$, $v$, $M_{\rm Ni}$, $\kappa_{\gamma}$),
 corresponding to the $^{56}$Ni-decay model;
 when the contributions from $^{56}$Ni and the magnetar are both important, the model can be termed as the hybrid model.
It should be noted that if the photospheric velocity $v_{\rm ph}$ is not measured, then $v \simeq v_{\rm ph}$ is also
 a free parameter.

\section{Fits to the light curves of SN~2010ay, SN~2006nx, and SN~14475}
\label{sec:fit}

To verify this unified model, we select three luminous SNe Ic-BL, SN~2010ay, SN~2006nx, and SN~14475.
SN~2010ay was discovered by the Catalina Real-time Transient Survey \citep[CRTS;][]{Dra2009} with
V-Band and R-Band peak absolute magnitudes, $M_V \approx -19.4$~mag and $M_R \approx -20.2 \pm 0.2$~mag, respectively.
We use the R$-$band light curve as a proxy for the bolometric light curve of SN~2010ay.
 Although the former is not a completely reliable proxy for the latter in some cases \citep[e.g.,][]{Wal2014},
 \citet{Whe2014} pointed out that, for SN~1993J (Type IIb), SN~1998bw (Ic-BL) and SN~2002ap (Ic-BL),
 the discrepancies between the R$-$band light curves and
 the quasi$-$bolometric light curves are rather small and can be neglected.

SN~2006nx and SN~14475 were discovered by the SDSS-II \citep{Tad2015} with
peak absolute bolometric magnitude $\approx-20.36$~mag and $\approx-20.1$~mag, respectively.
\citet{Tad2015} have constructed their bolometric light curves, so hereafter we adopt their data.

The peak luminosities of SN~2010ay, SN~2006nx, and SN~14475 are $\sim$~4.2, 4.84, and 3.8
 times that of SN~1998bw which is a well-studied SN Ic-BL with peak absolute magnitude $\sim-18.65$
  and the $^{56}$Ni mass $\sim0.43^{+0.05}_{-0.05}~M_\odot$ \citep{Maz2006}, respectively.
Hence, SN~2010ay, SN~2006nx, and SN~14475 are luminous SNe/HNe
 and brighter than all GRB-associated SNe/HNe confirmed to date.

The so-called Arnett law \citep{Arn1979,Arn1982}
 reports that the peak luminosity of a SN purely powered by $^{56}$Ni decay is proportional to the instantaneous energy deposition,
 therefore is proportional to the initial $^{56}$Ni mass \footnote{Strictly speaking, the peak luminosity of a SN depends sensitively upon
 the opacity, initial $^{56}$Ni mass, ejecta mass and kinetic energy. Only if the varieties other than $^{56}$Ni mass are same for SNe or
their effects on changing the peak luminosities of SNe can be canceled each other (e.g., higher
  opacity and lower ejecta mass, etc), then the
 peak luminosities of SNe are proportional to the initial $^{56}$Ni mass. These conditions could be satisfied in many SNe,
 so the Arnett law can be used to infer the $^{56}$Ni mass of SNe by comparing them to well-studies SNe (e.g., SN~1987A, etc) which
 have precise measurements of $^{56}$Ni masses and bolometric light curves. In Section \ref{subsec:ni-fit}, we will use the $^{56}$Ni-decay model to reproduce the
 light curves of SN~2010ay, SN~2006nx and SN~14475, and examine the $^{56}$Ni masses derived
 by the Arnett law.}, which is widely invoked to infer the $^{56}$Ni yields
 of some SNe \citep{Con2000,Str2002,Can2003,Str2005,Yuan2010,Cano2014}. Comparing
 to the type Ic supernova SN 1998bw, the $^{56}$Ni yields of SN~2010ay, SN~2006nx, and SN~14475 are estimated to be $1.81^{+0.21}_{-0.21}~M_\odot$,
 $2.08^{+0.24}_{-0.24}~M_\odot$, and $1.63^{+0.19}_{-0.19}~M_\odot$, respectively, when the Arnett law is applied.
The value $M_{\rm Ni}~\sim~1.81^{+0.21}_{-0.21}~M_\odot$ for SN~2010ay is considerably larger than the value $0.9^{+0.1}_{-0.1}~M_\odot$ derived in \citet{San2012}
while the above values of $M_{\rm Ni}$, $2.08^{+0.24}_{-0.24}~M_\odot$ for SN~2006nx and $1.63^{+0.19}_{-0.19}~M_\odot$ for SN~14475
 are roughly consistent with the values $M_{\rm Ni}=1.86^{+0.12}_{-0.12}~M_\odot$ and $M_{\rm Ni}=1.27^{+0.08}_{-0.09}~M_\odot$ for the two SNe derived in \citet{Tad2015}.

The inferred ejecta masses of SN~2010ay, SN~2006nx, and SN~14475
 are $M_{\rm ej}\approx 4.7~M_{\odot}$ \citep{San2012}, $7.52_{-1.74}^{+4.83}~M_\odot$ \citep{Tad2015},
 and $2.90_{-1.48}^{+3.38}~M_\odot$ \citep{Tad2015}, respectively.
Hence, the ratios of $M_{\rm Ni}$ to $M_{\rm ej}$ are $\sim0.4$, $\sim0.2$, and $\sim0.2-0.4$
for SN~2010ay, SN~2006nx, and SN~14475, respectively. These values far exceed the values of all GRB-associated SNe, which are typically $\sim0.05-0.1$ \citep{San2012} and larger than
the upper limit ($\sim 0.2$) for CCSNe \citep{Ume2008}.
The above analysis suggests that there should be other energy sources aiding these three SNe to the high peak-luminosities.
As mentioned above, power from ionized element re-combination and CSM-interaction can be safely neglected,
 we focus on the contributions of $^{56}$Ni and the magnetar possibly leaved behind the stellar explosion.

In this section, we use equations given in Section \ref{sec:uni} to
 reproduce the semi-analytical light curves and fit the observations of SN~2010ay, SN~2006nx, and SN~14475.
We assume that the fiducial value of the optical opacity $\kappa$ is 0.07 \citep[e.g.,][]{Tad2015}.
The photospheric expansion velocity $v_{\rm ph}$ of SN~2010ay, SN~2006nx, and SN~14475 are $\approx1.92 \times10^{4}$~km~s$^{-1}$ ($0.064~c$) \citep{San2012},
$\approx1.54 \times10^{4}$~km~s$^{-1}$ ($0.051~c$), and $\approx1.87 \times10^{4}$~km~s$^{-1}$  ($0.062~c$)  \citep{Tad2015}, respectively.
Thus $v$ is no longer a free parameter in our fitting.

In many light-curve modeling, $\kappa_{\gamma}$ has been
 assumed to have a fiducial
value 0.027 cm$^2$~g$^{-1}$ for the $^{56}$Ni-powered SNe Ic \citep[e.g.,][]{Cap1997,Maz2000,Mae2003} and $\gtrsim 0.01$ cm$^2$~g$^{-1}$
 for the magnetar-powered SNe \citep[see][Fig. 8]{Kot2013}, respectively.
It is worth emphasizing that in the magnetar model,
$\kappa_{\gamma}$ can vary between $\sim0.01$ and 0.2 cm$^2$~g$^{-1}$ ($E_{\gamma} \gtrsim 10^6$~eV) while $\kappa_{\rm X}$
  can vary between $\sim0.2$ and $10^4$ cm$^2$~g$^{-1}$ ($10^2$~eV $\lesssim E_{\rm X} \lesssim 10^6$~eV) \citep[see][Fig. 8]{Kot2013}.
Therefore, $\kappa_{\gamma}$ is also a free parameter in our fitting.

\subsection{The $^{56}$Ni-decay energy model}
\label{subsec:ni-fit}

\begin{figure}[htbp]
\begin{center}
   \includegraphics[width=0.48\textwidth,angle=0]{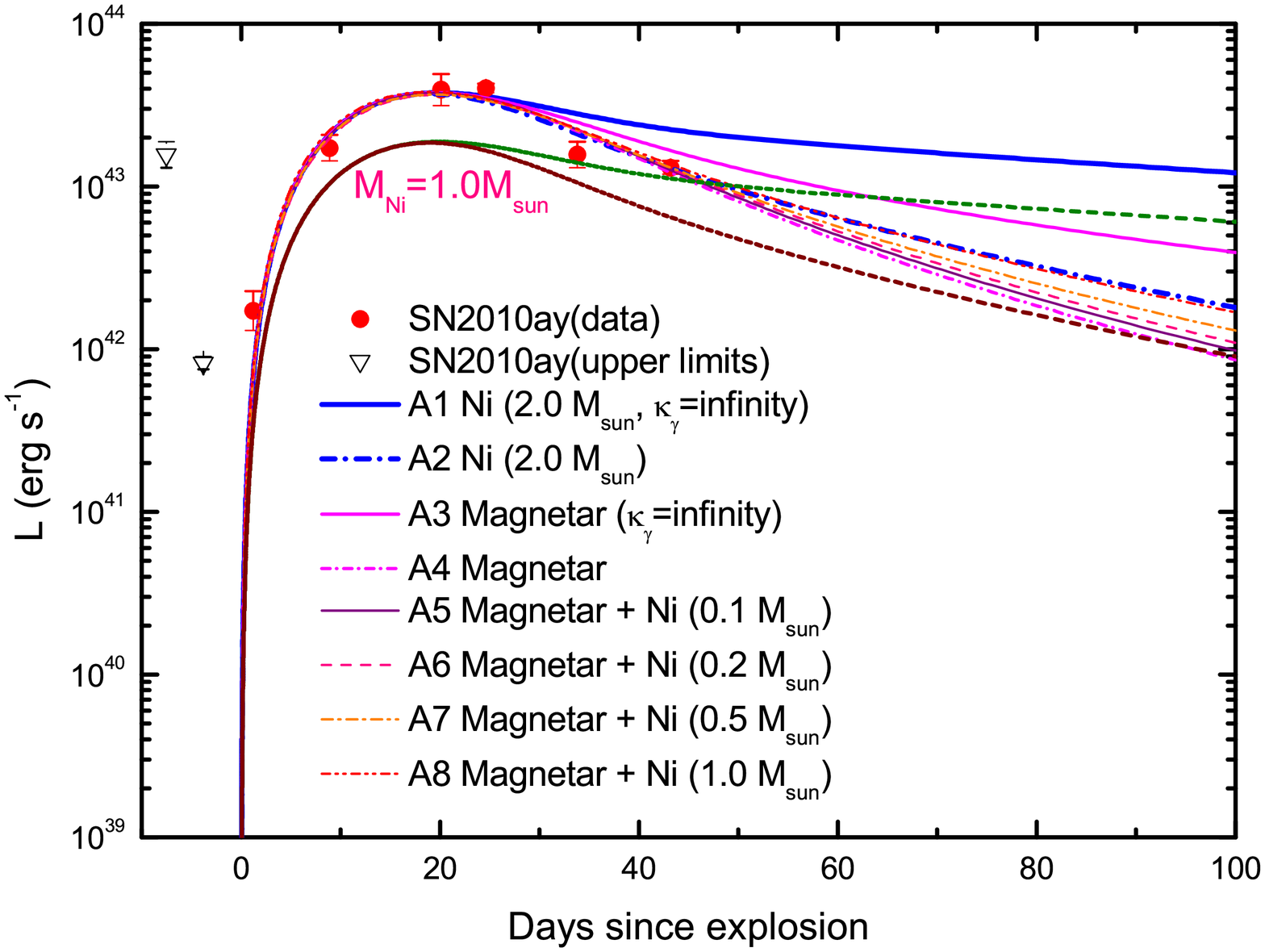}
   \includegraphics[width=0.48\textwidth,angle=0]{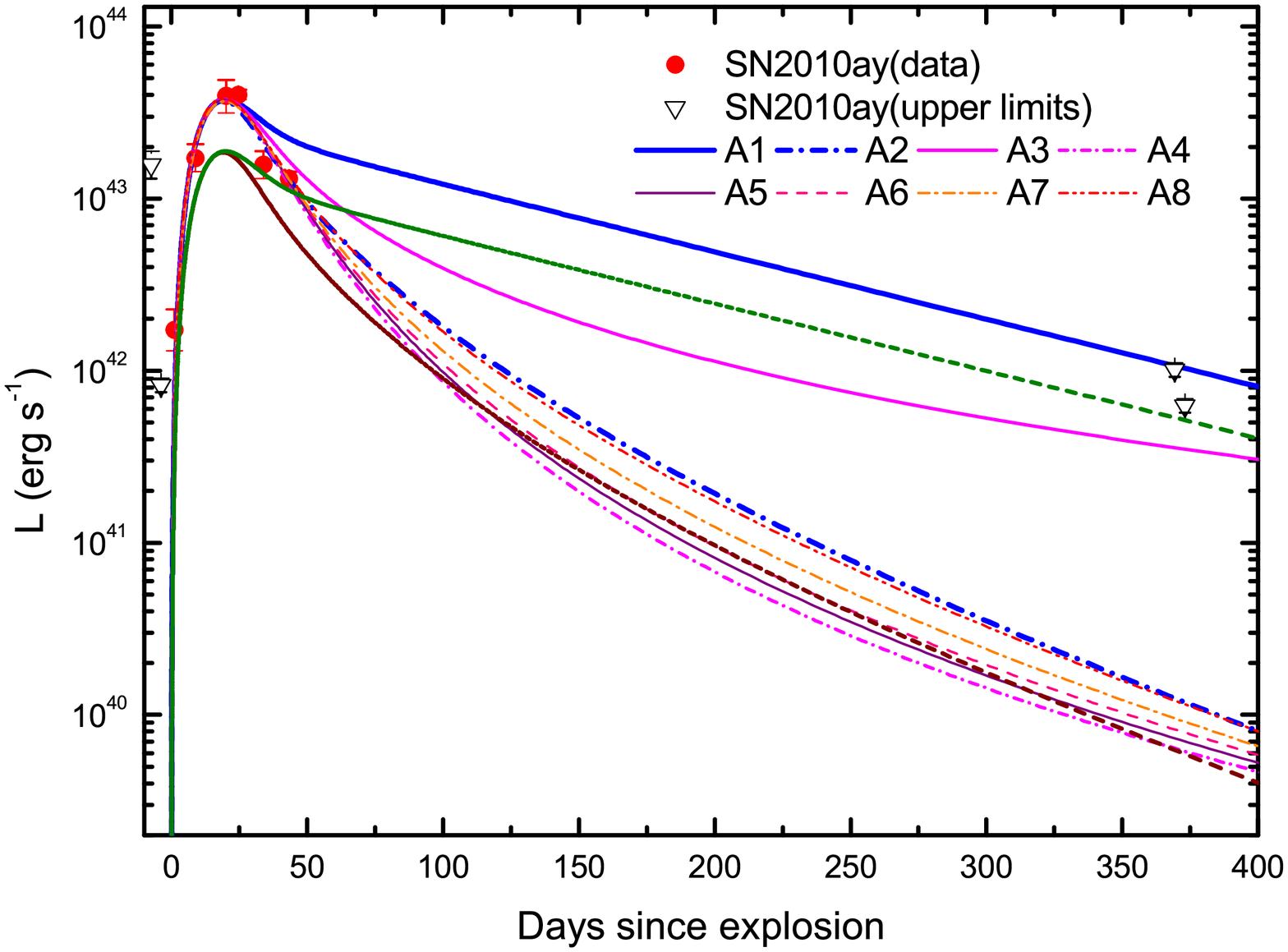}

   \includegraphics[width=0.48\textwidth,angle=0]{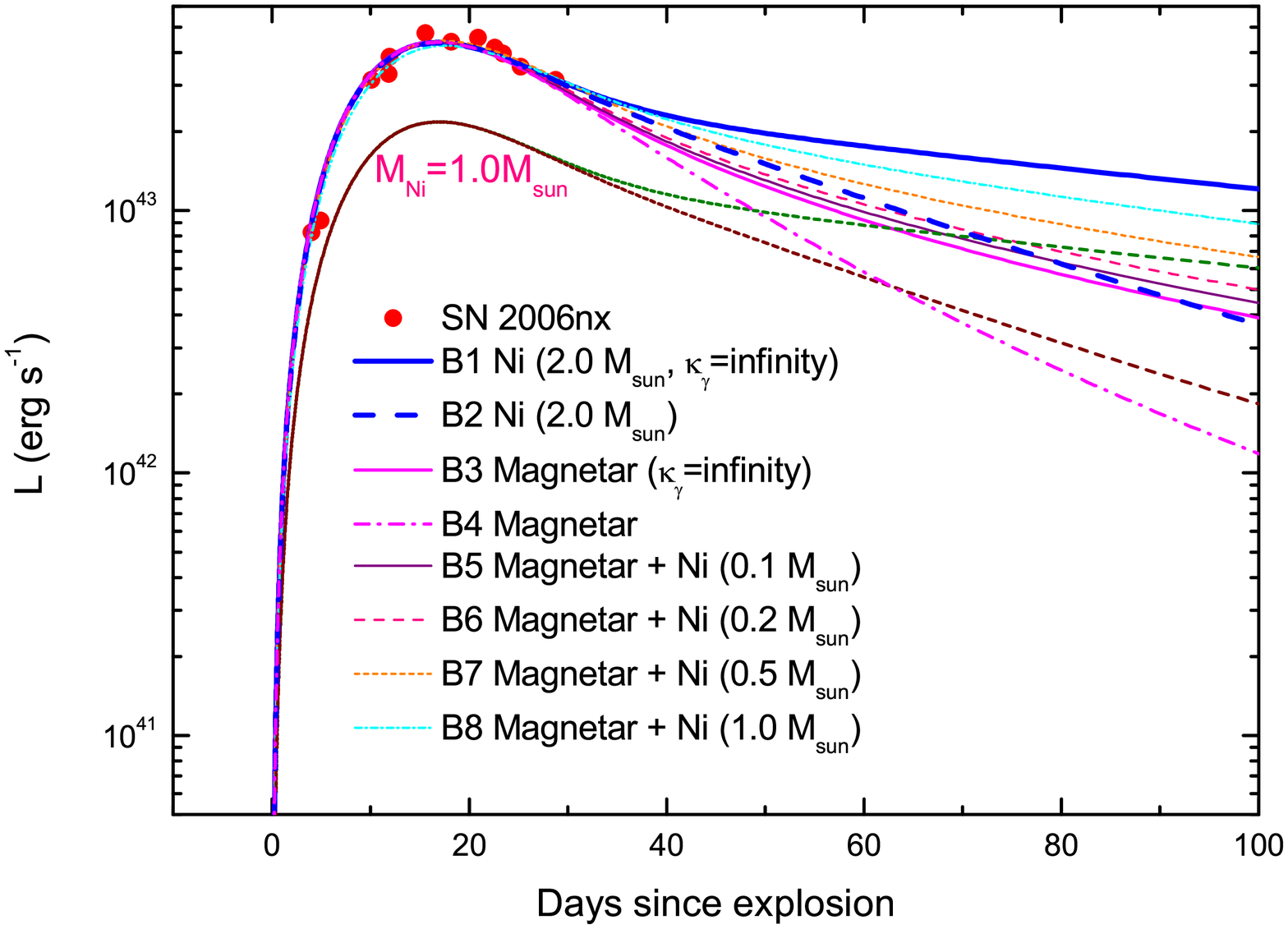}
   \includegraphics[width=0.48\textwidth,angle=0]{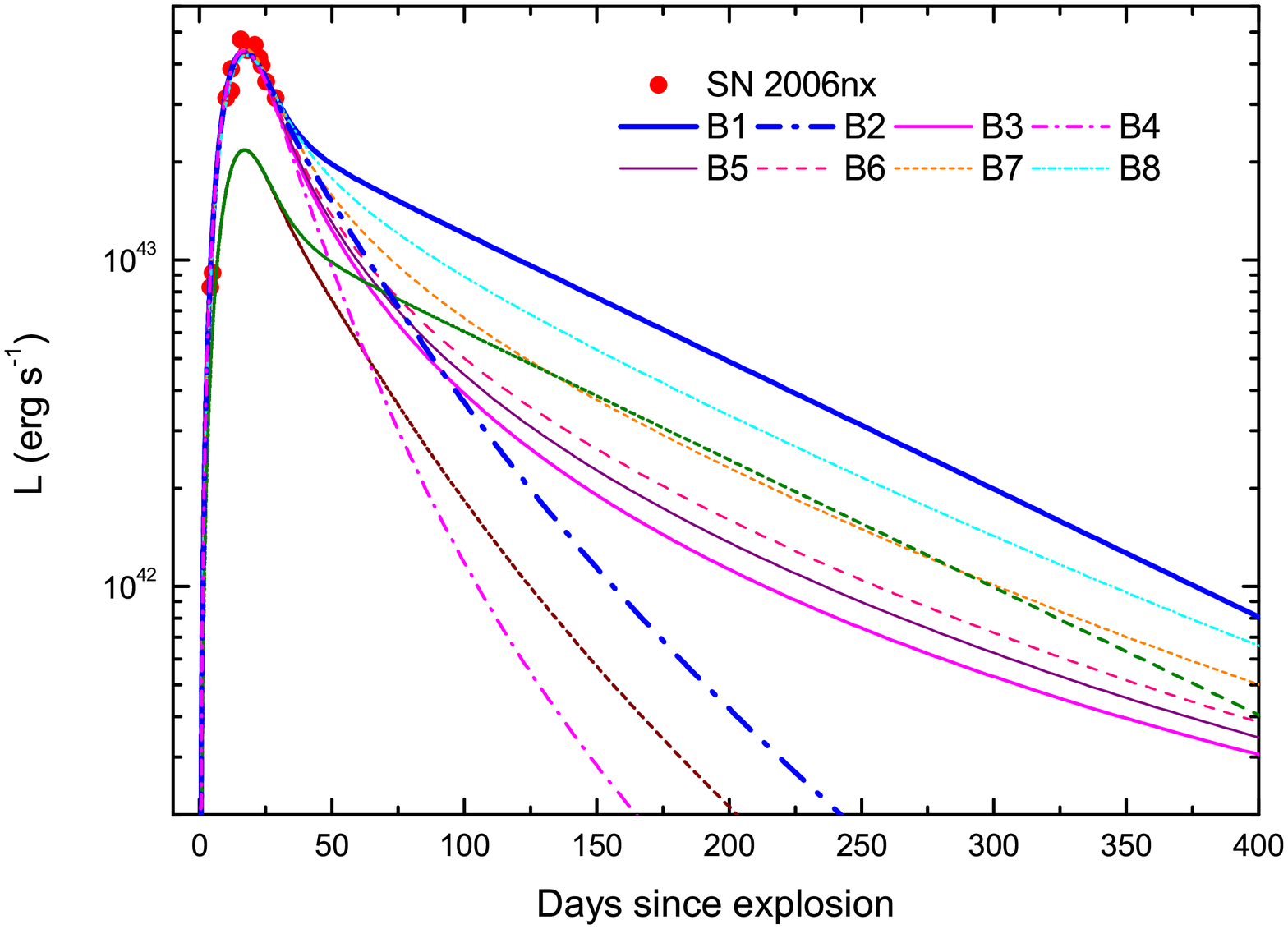}

   \includegraphics[width=0.48\textwidth,angle=0]{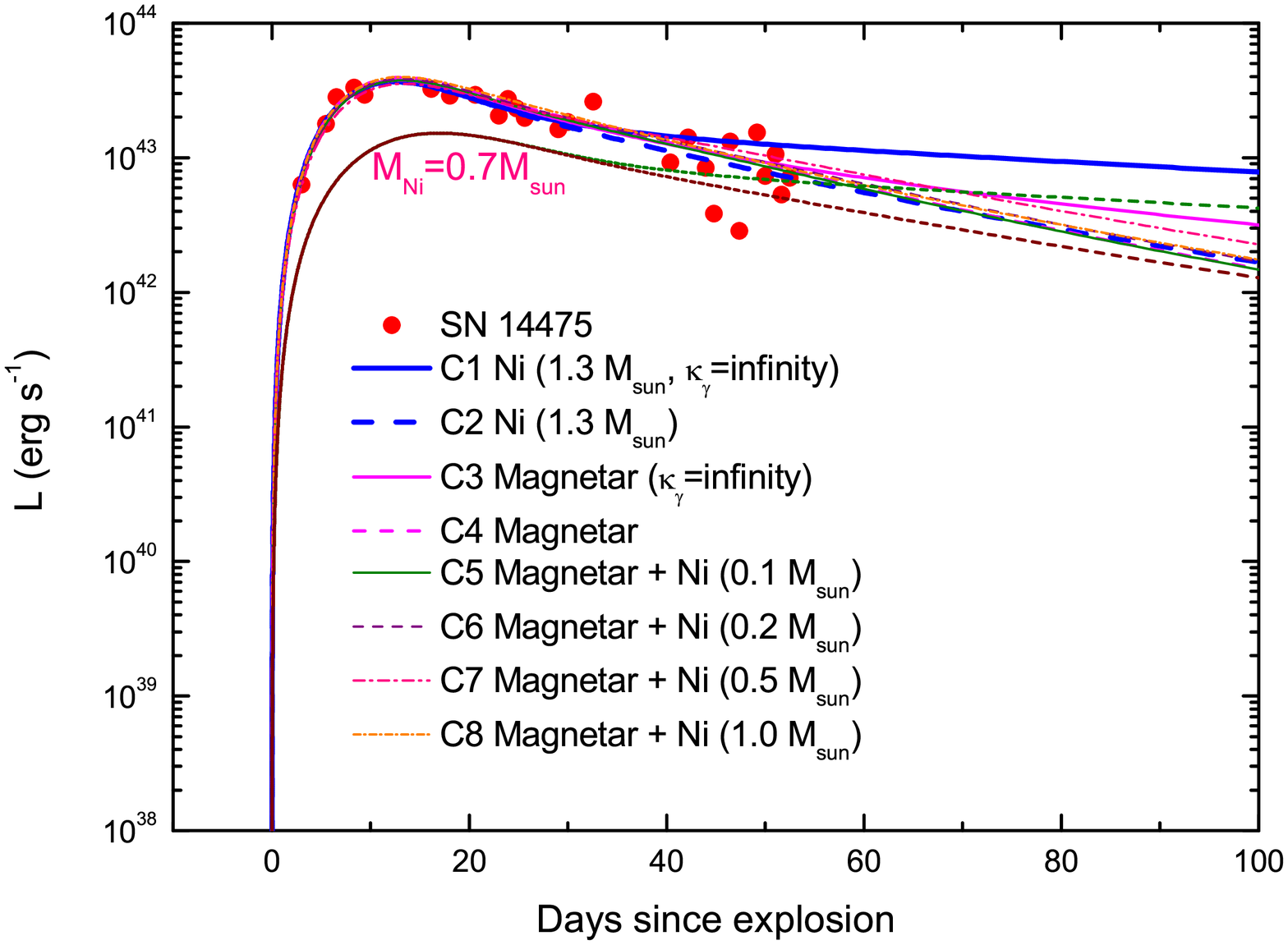}
   \includegraphics[width=0.48\textwidth,angle=0]{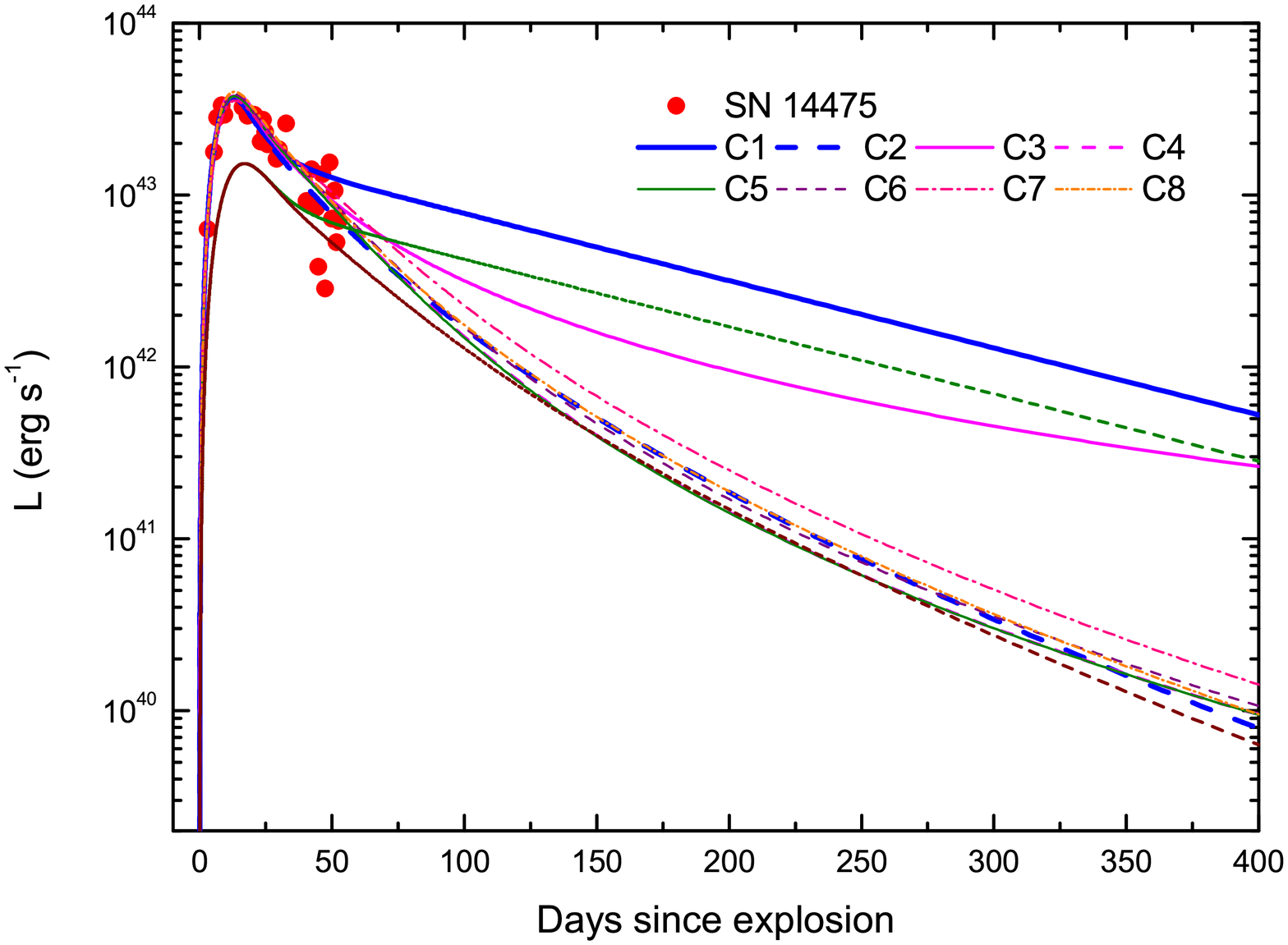}

   \caption{Model fits using the $^{56}$Ni-decay models, the magnetar models as well as the hybrid ($^{56}$Ni-decay + magnetar) models
    for SN~2010ay ($top$), SN~2006nx ($middle$), and SN~14475 ($bottom$).
    {\it Left panels}: -10$-$100~days after the explosion; {\it Right panels}: -10$-$400~days after the explosion.
   Data and upper limits of SN~2010ay are obtained from \citet{San2012}, data of SN~2006nx and SN~14475 are obtained from \citet{Tad2015}.
   The horizontal axis represents the time since the explosion in the rest frame.
   Parameters for Models A1-A8, B1-B8, and C1-C8 are shown in Table \ref{tab:para}. $\kappa$~=~0.07~cm$^2$~g$^{-1}$ is adopted.
   In all panels, we add two light curves reproduced by the
    $^{56}$Ni-decay models with 1~$M_{\odot}$ and 0.7~$M_{\odot}$, without and with
    $\gamma$-ray leakage. It can be seen from
    these light curves that 1~$M_{\odot}$ or 0.7~$M_{\odot}$ of $^{56}$Ni is inadequate to power these SNe.}
   \label{fig:fit}
\end{center}
\end{figure}

The parameters for the $^{56}$Ni-powered model are listed in Table \ref{tab:para} and the light curves
reproduced by these sets of parameters are shown
 in Fig. \ref{fig:fit} (Models A1 and A2 for SN~2010ay, B1 and B2 for SN~2006nx, C1 and C2 for SN~14475).

Using the empirical relation between
 the peak absolute magnitude and the mass of $^{56}$Ni derived by \citet{Dro2011}, \citet{San2012}
 inferred that the $^{56}$Ni mass of SN~2010ay is $~0.9_{-0.1}^{+0.1}~M_\odot$.
 However, it can be seen from Fig. \ref{fig:fit} and Table \ref{tab:para} that
$1.0~M_\odot$ of $^{56}$Ni is inadequate to power the peak luminosity of SN~2010ay solely.
 To power the peak luminosity, $\sim 2.0~M_\odot$ of $^{56}$Ni must be required.
The value $2.0~M_\odot$ is consistent with the value $1.81^{+0.21}_{-0.21}~M_\odot$ derived
 by the Arnett law.

It can be seen from Fig. \ref{fig:fit} and Table \ref{tab:para} that
$2.0~M_\odot$ and $1.3~M_\odot$ of $^{56}$Ni are required to power the peak luminosities of SN~2006nx and SN~14475, respectively.
The values $2.0~M_\odot$ and $1.3~M_\odot$ are consistent with the value $1.86^{+0.12}_{-0.12}~M_\odot$ and $1.27^{+0.08}_{-0.09}~M_\odot$
 derived by \citet{Tad2015} as well as the values $2.08^{+0.24}_{-0.24}~M_\odot$ and $1.63^{+0.19}_{-0.19}~M_\odot$ derived
 by the Arnett law.

In this paper, we adopt a fiducial value of optical opacity 0.07 cm$^2$~g$^{-1}$.
In many other papers, however, the fiducial values of opacity have also been assumed to be 0.06 cm$^2$~g$^{-1}$ \citep[e.g.,][]{Val2011,Lym2014}, 0.08 cm$^2$~g$^{-1}$ \citep[e.g.,][]{Arn1982,Maz2000}, 0.10 cm$^2$~g$^{-1}$ \citep[e.g.,][]{Nug2011,Inse2013,Whe2014} and 0.2 cm$^2$~g$^{-1}$ \citep[e.g.,][]{Kas2010,Nich2014}.
It is necessary to point out that more $^{56}$Ni for the $^{56}$Ni model or smaller $P_0$ of the magnetar for the magnetar model is required
 to account for the peak luminosity of SN~2010ay, SN~2006nx, and SN~14475
 when $\kappa$ is larger than 0.07 cm$^2$~g$^{-1}$ since larger
 $\kappa$ results in larger diffusion time and lower the peak luminosities.
 On the other hand, when $\kappa$ is smaller than 0.07 cm$^2$~g$^{-1}$, less
$^{56}$Ni or larger $P_0$ are required for the same peak luminosity. For example, If we assume that
 SN~2010ay is powered by $^{56}$Ni and $\kappa=$ 0.06 cm$^2$~g$^{-1}$,
 then $\sim 1.8~M_\odot$ of $^{56}$Ni must be synthesized, which is smaller than $2.0~M_\odot$ when $\kappa=$ 0.07 cm$^2$~g$^{-1}$.

To maintain the same peak luminosities and shapes of light curves, $\kappa M_{\rm ej}/v$ must be invariant.
 Because the photospheric velocities of SN~2010ay, SN~2006nx, and SN~14475 are fixed, so $\kappa M_{\rm ej}=$constant
 is required. Hence, smaller $\kappa$ requires larger $M_{\rm ej}$, as illustrated in Fig. \ref{fig:k-m}.

\subsection{The upper limits of $^{56}$Ni mass}
\label{subsec:limit}

A consensus has been reached that iron$-$core-collapse SNe can synthesize a large amount of $^{56}$Ni.
However, could these 3 SN explosions produce $\sim1.3~M_\odot$ or $\sim 2.0~M_\odot$ of $^{56}$Ni?
To get reasonable upper limits for the $^{56}$Ni masses synthesized in the explosions of SN~2010ay, SN~2006nx, and SN~14475,
we perform semi-quantitative estimates.

The radius of the $^{56}$Ni layer can be given as
\begin{eqnarray}
R_{\rm Ni}&=&\left(\frac{E_{\rm exp}+E_{\rm bin}}{\frac{4}{3}{\pi}aT^4}\right)^{1/3}\nonumber \\
  &\simeq & 4.0\times 10^3 \left(\frac{E_{\rm exp}+E_{\rm bin}}{10^{51}{\rm\; erg}}\right)^{1/3} \left(\frac{T}{5 \times 10^{9}{\rm\; K}}\right)^{-4/3}~\mbox{km}
\label{equ:Rni}
\end{eqnarray}
where $E_{\rm exp}$, $E_{\rm bin}$ are the explosion energy and the binding energy of the progenitor, respectively, $T$ is the
 temperature of the ejecta, $a~=~7.56566\times10^{15}$~erg~cm$^{-3}$~K$^{-4}$ is the radiation energy density constant.
Since $M_{\rm Ni}~={4/3}\rho_{\rm Ni}{\pi}R_{\rm Ni}^3$, $M_{\rm Ni}$ is therefore approximately proportional to
 ${E_{\rm exp}+E_{\rm bin}} \simeq {E_{\rm exp}}$
\footnote{\citet{Pej2014} demonstrated that in most cases the absolute value
of the progenitor binding energy $E_{\rm bin}$ must be less than the neutrino-driven wind energy $E_{\rm wind}$ (i.e., $E_{\rm exp}$) and the
 total energy $E_{\rm tot}$($={E_{\rm exp}+E_{\rm bin}}$) is approximately equal to the wind energy $E_{\rm wind}$,
 see their Fig. 12.}
 which in turn is approximately proportional to the kinetic energy $E_{\rm K}$, see, e.g., the bottom panel of Fig. 7 of \citet{Ham2003}
 and the left panel of Fig. 23 of \citet{Pej2014}.
The masses of the progenitors
 also influence the yields of $^{56}$Ni. These facts can also be seen in Fig. 7 and Table 3 in \citet{Ume2008}.
The typical masses of the $^{56}$Ni synthesized during the explosive silicon burning are $\sim 0.1-0.5 M_{\odot}$
for CCSNe.

We compare SN~2010ay, SN~2006nx, and SN~14475 with the prototype hypernova SN~1998bw.
The kinetic energy of SN~2010ay \citep[$\approx 1.1\times10^{52}$~erg,][]{San2012}
 and SN~14475 \citep[$\approx 4.71^{+12.29}_{-2.41}\times10^{51}$~erg,][]{Tad2015} are about a factor of 3-5 smaller
 than that of SN~1998bw \citep[$\approx 3-5\times10^{52}$~erg,][]{Iwa1998,Nak2001,Lym2014}, so we can infer that their $^{56}$Ni yields
 must be $\lesssim 0.5~M_\odot$ and $\sim 0.1-0.2~M_\odot$ which are significantly smaller than $1~M_\odot$ \footnote{ Although
  the kinetic energy of SN~14475 has rather significant uncertainty, its upper limit
  is only $\sim$ 1.7 $\times10^{52}$~erg, significant smaller than that of that of SN~1998bw, so we can
  expect that its $^{56}$Ni mass should be smaller that of SN~1998bw.}.
The kinetic energy of SN~2006nx \citep[$\approx 2.160^{+3.753}_{-0.501}\times10^{52}$~erg,][]{Tad2015} has rather significant uncertainty,
 but its upper limit ($\approx 5.91 \times10^{52}$~erg) is slightly larger than
 that of SN~1998bw ($\approx 5\times10^{52}$~erg). Hence,
 the $^{56}$Ni yield of SN~2006nx must be $\sim 0.5~M_\odot$ and smaller than $1~M_\odot$.

We can constrain the $^{56}$Ni masses of SN~2010ay, SN~2006nx, and SN~14475 using another method.
\citet{Ume2008} calculated the mass of the synthesized $^{56}$Ni in CCSNe with progenitor
 Zero-Age Main-Sequence masses $M_{\rm ZAMS}\leq 100M_{\odot}$ and found that the typical $^{56}$Ni mass is at most 20\% of the ejecta mass.
 According to our light-curve modeling, the ejecta mass of SN~2010ay, SN~2006nx, and SN~14475
 to be 6.5~$M_{\odot}$, 3.6~$M_{\odot}$, and 2.1 ~$M_{\odot}$, respectively, see Table \ref{tab:para}.
Combining these two results, $\lesssim$ 1.3, 0.7, and 0.4~$M_\odot$ of $^{56}$Ni can be synthesized in the explosions, respectively.
\footnote{It should be emphasized that these values are very rough and give rather loose upper limits. Nevertheless,
 real values of the masses of $^{56}$Ni are difficult to be larger than them and the range of $0.1-0.5~M_\odot$ is reasonable for these SNe.}
If the masses of $^{56}$Ni are $\sim$ 2.0, 2.0, and 1.3~$M_\odot$, the values of
 $M_{\rm Ni}/M_{\rm ej}$ are $\sim$ 0.31, 0.56, and 0.62 (see Table \ref{tab:para}) which (far) exceed the upper limit ($\sim 0.2$)
 of the mass ratio given by \citet{Ume2008} for CCSNe.

These results indicate
 that the $^{56}$Ni-decay model cannot account for the light curves of SN~2010ay, SN~2006nx, and SN~14475
 and these three SNe are probably solely or partly powered by other energy sources.

\subsection{The magnetar spin-down energy model}

Encountered by the above difficulty, alternative energy-reservoir models must be seriously considered.
According to their inferred ratio of $M_{\rm Ni}$ to $M_{\rm ej}$, $\sim0.2$, \citet{San2012} suggested
 that SN ejecta-circumstellar medium interaction could also contribute to the high peak
 luminosity of SN~2010ay as well, yet they did not perform further investigation for this
 possibility. Owing to the lack of narrow and/or intermediate-width emission
 lines indicative of interactions between the SN ejecta and hydrogen- and helium-deficient CSM,
 we argue that ejecta-CSM interactions could hardly provide a reasonable interpretation for the excess luminosities of
 SN~2010ay, SN~2006nx, and SN~14475. Hence we attribute the luminosity excess instead to energy injection by the nascent magnetars
 and employ the magnetar model to explain the data for these SNe Ic.

The parameters for the magnetar model are listed in Table \ref{tab:para} and the light curves
reproduced by the sets of parameters are shown in Fig. \ref{fig:fit} (Models A3 and A4 for SN~2010ay, B3 and B4 for SN~2006nx, C3 and C4 for SN~14475).
It can be seen from Fig. \ref{fig:fit} that
 the magnetar model can well fit the data.
 These results indicate that this model can be responsible for the light curves of SN~2010ay, SN~2006nx, and SN~14475,
  i.e., SN~2010ay, SN~2006nx, and SN~14475
  are probably powered by nascent millisecond (or $\sim$ 10 ms) magnetars.

\subsection{The hybrid ($^{56}$Ni + magnetar) model}

When the ejecta masses of CCSNe are not very large, e.g.,
 $\lesssim 2~M_\odot$, the $^{56}$Ni synthesized are
generally $\lesssim 0.1~M_\odot$, which contribute a minor fraction of
 the luminosity of luminous SNe and can be neglected in the fit.
\citet{Inse2013} demonstrated that the magnetar models
 without $^{56}$Ni and with $\sim0.1~M_\odot$ of $^{56}$Ni
 are similar and therefore neglected the contribution from
  $^{56}$Ni for their six SLSNe Ic.

However, SN~2010ay, SN~2006nx, and SN~14475 are not SLSNe and their ejected
 $^{56}$Ni might have a mass of $\sim 0.1-0.5 M_{\odot}$, so
 the contribution from $^{56}$Ni must be taken into account in our modeling.

Since the exact values of $M_{\rm Ni}$ of these three SNe are unknown
 when the contribution from magnetars is considered,
 we plot light curves corresponding to a variety of masses of $^{56}$Ni.
The typical values of $M_{\rm Ni}$ adopted here are 0.1, 0.2, 0.5
 and 1.0 $M_{\odot}$ \footnote{Although we have
 demonstrated in Section \ref{subsec:limit} that the
$^{56}$Ni mass of these three SNe can hardly be larger than
 0.5~$M_\odot$, considering the possible uncertainties,
 we also adopt the value of 1.0 $M_{\odot}$ in our modeling.}.
The parameters for the hybrid ($^{56}$Ni + magnetar) model
 are also listed in Table \ref{tab:para} and a family of light curves
reproduced by these sets of parameters are shown in Fig. \ref{fig:fit}
 (Models A5$-$A8 for SN~2010ay, B5$-$B8 for SN~2006nx, C5$-$C8 for SN~14475).

The $^{56}$Ni mass with $\lesssim 0.5~M_\odot$, the energy
 released by the rapidly spinning magnetar is
 still the dominant energy source in powering the
  optical light curves of SN~2010ay, SN~2006nx, and SN~14475.
When the $^{56}$Ni mass is $\sim 1~M_\odot$, the
 energy released by $^{56}$Ni is comparable with
or exceed the energy released by the magnetar.
If the $^{56}$Ni masses are $\lesssim 1~M_\odot$ and the magnetar parameters
 are adjusted, the light curves reproduced by the hybrid model
  are all in good agreement with observations,
 indicating that $\lesssim 1~M_\odot$ of $^{56}$Ni is admitted
 to be synthesized in their explosions.

On the other hand, there are rather significant discrepancies
 in the late-time ($\gtrsim 50$~days) light curves
 reproduced by different masses of $^{56}$Ni. These discrepancies
 cannot be eliminated by adjusting the magnetar parameters
 since the decline rate of late-time light curves is determined mainly by
 $^{56}$Co decay rate and the value of $\kappa_{\gamma}$.

We cannot determine the precise masses of $^{56}$Ni synthesized
 in the explosions of these SNe, because there is no precise late-time observation.
Provided that there are some late-time data observed,
 it would be powerful enough to constrain the precise
 value of their $^{56}$Ni mass.

To compare these models, we calculate the values of $\chi^2$/d.o.f for all these
 theoretical light curves \footnote{Since \citet{Tad2015} did not provide the observational errors of the data of SN~2006nx and SN~14475,
 we adopt a fiducial value for the error, $\sim1.3 \times 10^{42}$ erg s$^{-1}$, which is the smallest error of the SN~2010ay data.
 the difference between the real values of errors and our adopted values
 should not change the $\chi^2$/d.o.f significantly.}, see Table \ref{tab:para}.
 Since the $^{56}$Ni-powered models are disfavored
 in explaining these three SNe, as discussed above, we do not discuss them here even if they give smaller $\chi^2$/d.o.f..
 Therefore, we only consider the magnetar-powered models and hybrid models. It can be seen from the
 last column in Table \ref{tab:para} that Models A4 (magnetar-powered model with hard emission leakage),
 B7 (magnetar and 0.5~$M_\odot$ of $^{56}$Ni), and C5 (magnetar and 0.1~$M_\odot$ of $^{56}$Ni)
 have the smallest $\chi^2$/d.o.f. Nevertheless, due to the absence of late-time data,
 we still cannot conclude that these models are most favorable ones.

\subsection{An analysis for the origin of the kinetic energy}

It is necessary to take into account the kinetic energy coming from the $P$d$V$ work.
 The tapped rotational energy of a nascent magnetar
 ($E_{\rm NS}\simeq 2\times 10^{52}\left({P_0}/{1~{\rm ms}}\right)^{-2}$~erg)
 must be split into radiation energy ($E_{\rm rad}$) and
 kinetic energy ($E_{\rm K,mag}$).
 The former heat the ejecta, while the latter accelerate the ejecta.
When $P_0\sim9$~ms, $E_{\rm NS} \simeq 3.0~\times 10^{50}~{\rm erg}$.
 Even if all of the rotational energy of the putative magnetar
 is converted into kinetic energy of the ejecta \footnote{The calculations
 performed by \citet{Woos2010} have shown that $E_{\rm k,mag} \simeq 0.4~E_{\rm NS}$
 when $P_0~=~4.5$ ms, $B~=~1\times10^{14}$ G.},
 $E_{\rm K,mag} \approx E_{\rm NS} \simeq 3.0~\times 10^{50}~{\rm erg}$, it is still
  far less than the total kinetic energy
 ($E_{\rm K} \gtrsim 5\times 10^{51}$~erg) of SN~2010ay, SN~2006nx,
  as well SN~14475 and can therefore be completely neglected.

We now turn our attention to the neutrino-driven mechanism \citep{Bet1985}.
Hydrodynamic supernova simulations suggest that the neutrino-driven mechanism
 can produce SN energies $\lesssim 2\times10^{51}$~erg \citep{Ugl2012},
 far less than the observed kinetic energy of
 $E_{\rm K} \gtrsim 5 \times 10^{51}$~erg, indicating that
 there must be another mechanism accounting for the additional energy of $\gtrsim 3\times10^{51}$~erg.
A popular scenario to explain energetic SNe with kinetic energies
 $\gtrsim 10^{52}$~erg is the ``collapsar" model involving a ``BH-disk" system,
 in which the ``BH-disk" system generates bipolar jet/outflow while
 the ``disk wind" ensures that the progenitor star explodes and
 synthesizes enough amount of $^{56}$Ni.
However, our fittings have demonstrated that the newly formed magnetar
 did not collapse to a black hole at least several days after the explosion.
Hence, the huge kinetic energy might stem from
 some magnetohydrodynamic processes (e.g.,
 magnetic buoyancy, magnetic pressure, hoop stresses, etc)
 related to the proto-NS themselves \citep{Whe2000} after which the magnetars
 inject their rotational energy to heat
 the ejecta and generate the light curves.

\section{Discussion and Conclusions}
\label{sec:con}

Most type Ic CCSNe produce $^{56}$Ni which release energy to the ejecta via cascade decay and central NSs which convert
 a portion of their rotational energy into heating energy of the ejecta via the magnetic dipole radiation.
For almost all normal CCSNe, $\sim$ 0.1$-$0.7 $M_{\odot}$
 of $^{56}$Ni are adequate to power the peak and post-peak decline of observational light curves
 and the contribution of NSs can be neglected \footnote{If the explosion
 leaves a BH or nothing (corresponding to the so-called ``pair instability SNe" \citep{Heg2002} rather than CCSNe),
 then no NS contributes luminosity to the SN and the
 SN is solely powered by $^{56}$Ni; if the explosion produces a transitional NS lasting about several hours or days
 the contribution of the NS to the light curve of the SN can also be neglected.}.
 In contrast, for SLSNe Ic which are lack of evidence of interaction
 and cannot be explained by $^{56}$Ni solely, millisecond NSs (magnetars)
 can supply almost all input power and the contribution of $^{56}$Ni can be neglected.

For luminous SNe Ic with $-19.5\gtrsim M_{\rm peak}\gtrsim-21$~mag in any band, however,
 both contributions from $^{56}$Ni and NSs cannot be neglected without detailed modeling since the $^{56}$Ni permitted
 by completely explosive burning are usually
 inadequate to power the high peak luminosities while the production of $\sim0.1-\sim0.5 M_{\odot}$
 of $^{56}$Ni is reasonable and may contribute a significant portion of the total luminosity.
 Therefore, these ``gap-filler" events which bridge normal SNe and SLSNe must be
 explained by a unified model which contain contributions from $^{56}$Ni and NSs.

To illustrate the necessity of constructing the unified model,
 we select and apply this model to three luminous SNe Ic-BL,
SN~2010ay, SN~2006nx, and SN~14475.
We tested the possibility that the luminosity evolutions of SN~2010ay, SN~2006nx, and SN~14475
 are solely governed by radioactive $^{56}$Ni cascade decay using
 the semi-analytical method and demonstrated that too large amounts, roughly 2.0, 2.0, and 1.3~$M_\odot$ of $^{56}$Ni, are needed to
power their peak luminosities, respectively.
These results disfavor the $^{56}$Ni model in explaining the light curves
  of SN~2010ay, SN~2006nx, and SN~14475, since the ratios of
 $M_{\rm Ni}$ to $M_{\rm ej}$ are 0.31, 0.56, and 0.62, respectively,
  significantly exceeding the upper limit ($\sim0.2$) calculated for CCSNe \citep{Ume2008}.

Then we alternatively invoke the magnetar model to reproduce the
 light curves of SN~2010ay, SN~2006nx, and SN~14475
 and show that the light curves reproduced
 can be well consistent with the observations, suggesting that SN~2010ay, SN~2006nx, and SN~14475
 are probably powered by spin-down energy injection
 from newly born millisecond magnetars.
It seems that the magnetar model can account for the light curves of these SNe and $^{56}$Ni
 is not necessary to be introduced. However, a moderate amount of $^{56}$Ni, $\sim0.1-0.5 M_{\odot}$,
 can be expected in completely explosive burning of silicon shell
  and could give rather energy to those not very luminous SNe.
So $^{56}$Ni cannot be neglected when the SNe are not very luminous.
 Thus we use the hybrid model to fit these three SNe.

We set $M_{\rm Ni}$ = 0.1, 0.2, 0.5 and 1.0 $M_{\odot}$ and find that
 $\lesssim 0.2~M_\odot$ of $^{56}$Ni can not significantly influence the
theoretical light curves while $\gtrsim 0.5~M_\odot$ of $^{56}$Ni can
 effectively alters the shapes of the light curves, leading us to adjust the
magnetar parameters to fit the observations.
Therefore, all these models ($^{56}$Ni, magnetar, and hybrid), with
 different amount of $^{56}$Ni (from 0 to 2 or 1.3~$M_{\odot}$),
 fit the light curves equally well and can be unified into a unified model. Among these models,
 the $^{56}$Ni model needs huge, unrealistic amounts of $^{56}$Ni, resulting in over-high ratios of $M_{\rm Ni}$ to $M_{\rm ej}$.
 In contrast, the magnetar model needs no $^{56}$Ni and conflict with the basic theory of nuclear synthesis.
 The hybrid model is reasonable in explaining these luminous SNe.

These facts suggest that these three luminous SNe Ic are all powered by both $^{56}$Ni and magnetars, indicating that
 the magnetars contribute a large fraction of but not all power in driving their light curves.
So we can conclude that, when $M_{\rm peak}$ (in any band) is $\lesssim -21$~mag, the contribution of $^{56}$Ni can be neglected in light-curve modeling;
 when $-21 \lesssim M_{\rm peak}\lesssim -19.5$~mag, the contribution from the magnetar cannot be omitted in modeling while
 the contribution from $^{56}$Ni must be considered seriously;
 but when $M_{\rm peak} \gtrsim -19.5$~mag, the contribution of the magnetar can generally
 be neglected \citep[but see][]{Mae2007}.

It should be pointed out that the ``fiducial cutoffs" $-19.5$~mag and $-21$~mag that divide the normal SNe, luminous SNe and SLSNe
 are artificially given and do not have unambiguous physical meanings.
 Although no consistent picture of these SNe Ic with rather different peak luminosities has emerged so far,
 it seems that they have similar physical nature and similar power sources,
 e.g., $^{56}$Ni and newly born NSs (magnetars),
 while the former result in a relative narrow range of peak-luminosities
 owing to the existence of the upper limit of $^{56}$Ni yields
 and thus the peak luminosities, the latter
might produce a wide range of peak-luminosities since the period of newly born NSs
 can vary from several milliseconds to several seconds, and magnetic field strength of them can be up to $10^{15}$~G.
 These two power sources could result in a continuous sequence of
 peak-luminosities, covering normal, luminous, and very luminous SNe Ic and unifying most of them into a whole category
  in which normal SNe are powered by slow-spinning NSs and $^{56}$Ni, while luminous SNe and SLSNe
  are mainly powered by fast-spinning (millisecond) magnetar and $^{56}$Ni. The similarity between late-time spectra of
  SLSNe Ic and spectra of normal SNe Ic-BL \citep{Pas2010,Gal2012,Inse2013} also supports this unified picture.

Furthermore, we suggest that
 many SNe Ic with $M_{\rm peak} \lesssim -19.5$~mag might be partly powered by a nascent magnetar,
  i.e., the very young SN remnants of the explosions may harbor magnetars. The discrepancy in brightness
 between luminous SNe Ic and SLSNe Ic stem mainly from the initial spin periods of the magnetars,
 the former have $P_0~\sim 7-15$~ms, while the latter have $P_0~\sim 1-6$~ms \citep{Inse2013,Nich2014}.
 For luminous SNe Ic, power from magnetars with $P_0~\sim 10$~ms
 can become comparable to or exceed that from $^{56}$Ni, and vice versa;
 for SLSNe Ic, millisecond magnetars usually overwhelm $^{56}$Ni in powering them.

For the sake of completeness, both these two power sources
 must be taken into account in light-curve modeling.
Nevertheless, we still emphasize that this unified scenario is especially important for the gap-filler SNe Ic since
 they, as aforementioned, are hardly powered solely by $^{56}$Ni but $^{56}$Ni
 can contribute a non-negligible portion of the luminosity and must be considered in modeling.
 Modeling for these luminous SNe can also help us to understand
 how normal SNe Ic relate to SLSNe Ic.
 Besides, even if a SN or superluminous SN that can be solely explained by the magnetar model, Fe line in its spectrum, if observed,
 must be explained by introducing a moderate amount of $^{56}$Ni, because the magnetar model cannot explain Fe line \citep{Kas2010}.
This is another advantage of the unified model.
In the recent excellent review for SLSNe, \citet{Gal2012} already noticed these ``gap-filler" SNe and
 reckoned that they are likely to be intermediate events between radioactivity-powered SNe/SLSNe Ic and the SLSNe Ic powered by
 some other processes. Our analysis has confirmed this conjecture and furthermore demonstrated that these ``intermediate events" are
  likely to have the same power sources and the difference between them and their lower-luminosity and higher-luminosity cousins
 are mainly due to the physical properties of the nascent neutron stars embedded in their debris.

While this unified energy-reservoir model seems reasonable for SN~2010ay, SN~2006nx, and SN~14475,
 the precise value of $^{56}$Ni yields cannot be determined, indicating
that the traditional method that suppose all power given by $^{56}$Ni cascade decay
 is not always reliable in inferring the $^{56}$Ni yields of CCSNe, especially for those luminous and very luminous ones.
To discriminate these models and roughly determine the $^{56}$Ni mass as well as
 the parameters of the putative NSs, late-time ($\gtrsim 100-200$~d) data are required. Unfortunately, these
 three SNe are all lacking late-time observations, especially for SN~14475 whose luminosity measurements in
 the period between 25 and 35 days after the peak are rather uncertain.

\begin{figure}[htbp]
\begin{center}
   \includegraphics[width=1.0\textwidth,angle=0]{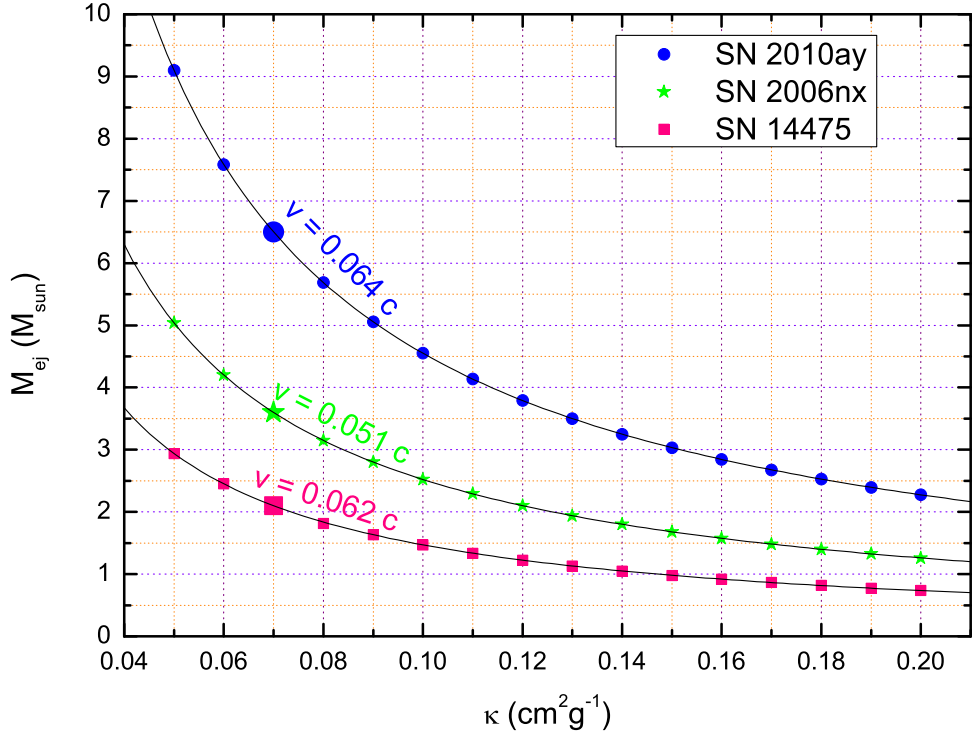}
   \caption{The values of $\kappa$ vs. the values of $M_{\rm ej}$ for SN~2010ay, SN~2006nx, and SN~14475.
   The values of $\kappa$ vary from 0.05~cm$^2$~g$^{-1}$ to 0.20~cm$^2$~g$^{-1}$, resulting in rather
   different values of $M_{\rm ej}$ required. Big Circle, Star, and Square correspond to SN~2010ay, SN~2006nx, and SN~14475 with $\kappa=0.07$~cm$^2$~g$^{-1}$, respectively. }
   \label{fig:k-m}
\end{center}
\end{figure}

Spectral analysis would be of outstanding importance in determining the precise values of $^{56}$Ni and $M_{\rm ej}$ as well as
 the parameters of the NS parameters.
In practice, detection and analysis of nebular emission from $^{56}$Fe resulted from the $^{56}$Ni cascade decay
 and other heavy elements are necessary for precise measurement of $^{56}$Ni yields.
The ejecta mass $M_{\rm ej}$ can significantly influence the rise time, shape, and peak luminosity as well
 as the ratio of $M_{\rm Ni}$ to $M_{\rm ej}$ of a SN, so it is also a very important quantity in modeling.
 Due to the uncertainty of optical opacity $\kappa$, $M_{\rm ej}$ cannot be precisely determined using solely the light-curve modeling
 (see Fig.\ref{fig:k-m}).
 To get more precise value of the ejecta mass or at least a rigorous upper/lower limit,
 as performed by \citet{Gal2009} for SN~2007bi,
 nebular modeling are required.

The above procedures need very precise observations and spectral analysis lasting several hundred days after the detection of the first light.
Unfortunately, no late-time photometric data and spectra have been observed for these three SNe, especially for SN~2006nx and SN~14475.
 We cannot get enough information to
 determine the precise masses of $^{56}$Ni and the real values of magnetar parameters.
 Hence, we cannot distinguish among these model parameters and therefore to have a more precise scenario.
We can expect that future high cadence multi-epoch UV$-$optical$-$NIR observations of the luminous SNe
 can pose stringent constraints on the parameters of the nascent NSs and $^{56}$Ni yields.

\begin{acknowledgements}
\acknowledgments

We thank an anonymous referee for helpful comments and suggestions that have allowed us to improve this manuscript.
We thank Xiangyu Wang, Yongfeng Huang, Xiaofeng Wang, Yongbo Yu, Jinjun Geng, Liangduan Liu and Junjie Wei for helpful discussion and comments.
This work is supported by the National Basic Research Program (``973" Program)
of China (grant Nos. 2014CB845800 and 2013CB834900) and the National Natural Science Foundation of China (grant Nos. 11033002 and 11322328). X.F.W was also partially supported by the One-Hundred-Talent Program, the Youth Innovation Promotion Association, and the Strategic Priority Research Program
``The Emergence of Cosmological Structures" (grant No. XDB09000000) of
the Chinese Academy of Sciences, and the Natural Science Foundation of Jiangsu Province (No. BK2012890).
\end{acknowledgements}


\clearpage

\begin{table*}
\begin{center}
\caption{Parameters for unified ($^{56}$Ni-decay, magnetar spin-down, and hybrid ($^{56}$Ni + magnetar)) modeling}
\label{tab:para}
\begin{tabular}{cccccccccccc}
\hline
\hline
 & $M_{\rm ej}$   & $B$           & $P$    & $v$    & $\kappa$ & $\kappa_{\gamma}$      & $M_{\rm Ni}$   & $M_{\rm Ni}/M_{\rm ej}$ & $\chi^2$/d.o.f \\
 & ($M_{\odot}$)  & ($10^{14}$~G) & (ms)   & ($c$)  & (cm$^2$ g$^{-1}$)  & (cm$^2$ g$^{-1}$)      & ($M_{\odot}$)  &     &     \\
\hline
\hline
{\bf SN~2010ay}\\
\hline
Model A1 $^{a}$ & 6.5 & 0    & $\infty $ & 0.064 & 0.07  & $\infty $   & 2.0  & 0.308 & 5.41\\
Model A2 $^{a}$ & 6.5 & 0    & $\infty $ & 0.064 & 0.07  & $ 0.014$    & 2.0  & 0.308 & 1.05\\
Model A3 $^{b}$ & 6.5 & 4.5  &    8.8    & 0.064 & 0.07  & $\infty $   & 0    & 0     & 2.36\\
Model A4 $^{b}$ & 6.5 & 4.5  &    8.8    & 0.064 & 0.07  & $ 0.021$    & 0    & 0     & 1.34\\
Model A5 $^{c}$ & 6.5 & 4.5  &    9.0    & 0.064 & 0.07  & $ 0.021$    & 0.1  & 0.015 & 2.00\\
Model A6 $^{c}$ & 6.5 & 4.5  &    9.3    & 0.064 & 0.07  & $ 0.021$    & 0.2  & 0.031 & 2.05\\
Model A7 $^{c}$ & 6.5 & 5.2  &    10.3   & 0.064 & 0.07  & $ 0.021$    & 0.5  & 0.077 & 2.11\\
Model A8 $^{c}$ & 6.5 & 7.2  &    12.5   & 0.064 & 0.07  & $ 0.021$    & 1.0  & 0.154 & 2.12\\
\hline
\hline

{\bf SN~2006nx}\\
\hline
Model B1 $^{a}$ & 3.6 & 0    & $\infty $  & 0.051 & 0.07  & $\infty $  & 2.0 & 0.556 &  1.24\\
Model B2 $^{a}$ & 3.6 & 0    & $\infty $  & 0.051 & 0.07  & $ 0.037$   & 2.0 & 0.556 &  1.22\\
Model B3 $^{b}$ & 3.6 & 4.5  &    8.8     & 0.051 & 0.07  & $\infty $  & 0   & 0     &  1.58\\
Model B4 $^{b}$ & 3.6 & 4.5  &    8.8     & 0.051 & 0.07  & $ 0.037 $  & 0   & 0     &  1.56\\
Model B5 $^{c}$ & 3.6 & 4.5  &    9.0     & 0.051 & 0.07  & $\infty $  & 0.1 & 0.028 &  1.60\\
Model B6 $^{c}$ & 3.6 & 4.5  &    9.2     & 0.051 & 0.07  & $\infty $  & 0.2 & 0.056 &  1.44\\
Model B7 $^{c}$ & 3.6 & 4.5  &    9.8     & 0.051 & 0.07  & $\infty $  & 0.5 & 0.139 &  1.11\\
Model B8 $^{c}$ & 3.6 & 4.8  &    11.9    & 0.051 & 0.07  & $\infty $  & 1.0 & 0.278 &  1.60\\

\hline
\hline

{\bf SN~14475}\\
\hline
Model C1 $^{a}$ & 2.1 & 0    & $\infty $  & 0.062 & 0.07  & $\infty $  & 1.3 & 0.619 &  4.10\\
Model C2 $^{a}$ & 2.1 & 0    & $\infty $  & 0.062 & 0.07  & $ 0.063$   & 1.3 & 0.619 &  2.78\\
Model C3 $^{b}$ & 2.1 & 4.8  &    10.4    & 0.062 & 0.07  & $\infty $  & 0   & 0     &  3.03\\
Model C4 $^{b}$ & 2.1 & 4.8  &    10.4    & 0.062 & 0.07  & $ 0.133$   & 0   & 0     &  3.30\\
Model C5 $^{c}$ & 2.1 & 4.8  &    10.7    & 0.062 & 0.07  & $ 0.133$   & 0.1 & 0.048 &  2.97\\
Model C6 $^{c}$ & 2.1 & 4.8  &    10.9    & 0.062 & 0.07  & $ 0.133$   & 0.2 & 0.095 &  3.53\\
Model C7 $^{c}$ & 2.1 & 4.8  &    12.6    & 0.062 & 0.07  & $ 0.133$   & 0.5 & 0.238 &  3.54\\
Model C8 $^{c}$ & 2.1 & 4.8  &    15.2    & 0.062 & 0.07  & $ 0.063$   & 1.0 & 0.476 &  4.03\\
\hline
\hline
\end{tabular}
\end{center}
$^{a}$ The $^{56}$Ni-decay model. \\
$^{b}$ The magnetar model. \\
$^{c}$ The hybrid ($^{56}$Ni + magnetar) model. \\
\end{table*}

\clearpage

\clearpage

\begin{thebibliography}{}

\bibitem[Arnett(1979)]{Arn1979}{Arnett}, W.~D. 1979, ApJL, 230, L37
\bibitem[Arnett(1982)]{Arn1982}{Arnett}, W.~D.\ 1982, \apj, 253, 785
\bibitem[Ben-Ami et~al.(2014)]{Ben-A2014}{Ben-Ami}, S., {Gal-Yam}, A., {Mazzali}, P.~A., {et~al.} 2014, \apj, 785, 37
\bibitem[Bethe \& Wilson(1985)]{Bet1985}{Bethe}, H.~A., \& {Wilson}, J.~R. 1985, \apj, 295, 14
\bibitem[Bucciantini et~al.(2008)]{Buc2008}{Bucciantini}, N., {Quataert}, E., {Arons}, J., {Metzger}, B.~D., \& {Thompson}, T.~A. 2008, \mnras, 383, L25
\bibitem[Cano et~al.(2014)]{Cano2014}{Cano}, Z., {de Ugarte Postigo}, A., {Pozanenko}, A., {et~al.} 2014, \aap, 568, 19
\bibitem[Cappellaro et~al.(1997)]{Cap1997}{Cappellaro}, E., {Mazzali}, P.~A., {Benetti}, S., {et~al.} 1997, \aap, 328, 203
\bibitem[Candia et al.(2003)]{Can2003}{Candia}, P., {Krisciunas}, K., {Suntzeff}, N.~B., {et~al.} 2003, \pasp, 115, 277
\bibitem[Chatzopoulos et~al.(2009)]{Cha2009}{Chatzopoulos}, E., {Wheeler}, J.~C., \& {Vinko}, J. 2009, \apj, 704, 1251
\bibitem[Chatzopoulos et~al.(2012)]{Cha2012}{Chatzopoulos}, E., {Wheeler}, J.~C., \& {Vinko}, J. 2012, \apj, 746, 121
\bibitem[Chatzopoulos et~al.(2013)]{Cha2013}{Chatzopoulos}, E., {Wheeler}, J.~C., {Vinko}, J., {Horvath}, Z.~L., \& {Nagy}, A. 2013, \apj, 773, 76
\bibitem[Chevalier(1982)]{Che1982}{Chevalier}, R.~A. 1982, \apj, 258, 790
\bibitem[Chevalier \& Fransson(1994)]{Che1994}{Chevalier}, R.~A., \& {Fransson}, C. 1994, \apj, 420, 268
\bibitem[Chevalier \& Irwin(2011)]{Che2011}{Chevalier}, R.~A., \& {Irwin}, C.~M. 2011, ApJL, 729, L6
\bibitem[Chugai(1994)]{Chu1994}{Chugai}, N.~N. 1994, \mnras, 326, 1448
\bibitem[Clocchiatti \& Wheeler(1997)]{Clo1997}{Clocchiatti}, A., \& {Wheeler}, J.~C. 1997, \apj, 491, 375
\bibitem[Contardo et al.(2000)]{Con2000}{Contardo}, G., {Leibundgut}, B., \& {Vacca}, W.~D. 2000, \aap, 359, 876
\bibitem[Colgate \& McKee(1969)]{Col1969}{Colgate}, S.~A., \& {McKee}, C. 1969, \apj, 157, 623
\bibitem[Colgate et~al.(1980)]{Col1980}{Colgate}, S.~A., {Petschek}, A.~G., \& {Kriese}, J.~T. 1980, ApJL, 237, L81
\bibitem[Dai \& Lu(1998a)]{Dai1998a}{Dai}, Z.~G., \& {Lu}, T. 1998a, \aap, 333, L87
\bibitem[Dai \& Lu(1998b)]{Dai1998b}{Dai}, Z.~G., \& {Lu}, T. 1998b, PhRvL, 81, 4301
\bibitem[Dai(2004)]{Dai2004}{Dai}, Z.~G. 2004, \apj, 606, 1000
\bibitem[Dai \& Liu(2012)]{Dai2012}{Dai}, Z.~G., \& {Liu}, R.~Y. 2012, \apj, 759, 58
\bibitem[Drake et~al.(2009)]{Dra2009}{Drake}, A.~J., {Djorgovski}, S.~G., {Mahabal}, A., {et~al.} 2009, \apj, 696, 870
\bibitem[Drout et~al.(2011)]{Dro2011}{Drout}, M.~R., {Soderberg}, A.~M., {Gal-Yam}, A., {et~al.} 2011, \apj, 741, 97
\bibitem[Drout et~al.(2013)]{Dro2013}{Drout}, M.~R. , {Soderberg}, A.~M., {Mazzali} P.~A., {et~al.} 2013, \apj, 774, 58
\bibitem[Falk \& Arnett(1977)]{Fal1977}{Falk}, S.~W., \& {Arnett}, W.~D. 1977, \apjs, 33, 515	
\bibitem[Filippenko(1997)]{Fil1997}{Filippenko}, A.~V. 1997, \araa, 35, 309
\bibitem[Gal-Yam et~al.(2009)]{Gal2009}{Gal-Yam}, A., {Mazzali}, P., {Ofek}, E. O., {et~al.} 2009, Natur, 462, 624
\bibitem[Gal-Yam(2012)]{Gal2012}{Gal-Yam}, A. 2012, Science, 337, 927
\bibitem[Ginzburg \& Balberg(2012)]{Gin2012}{Ginzburg}, S., \& {Balberg}, S. 2012, \apj, 757, 178
\bibitem[Hamuy(2003)]{Ham2003}{Hamuy}, M. 2003, \apj, 582, 905
\bibitem[Heger \& Woosley(2002)]{Heg2002}{Heger}, A., \& {Woosley}, S.~E. 2002, \apj, 567, 532
\bibitem[Hjorth \& Bloom(2012)]{Hjo2012}{Hjorth}, J., \& {Bloom}, J.~S., 2012, in Chapter 9 in Gamma-Ray Bursts,
 ed. C. Kouveliotou, R. A. M. J. Wijers, \& S. Woosley (Cambridge Astrophysics Series, Vol. 51; Cambridge: Cambridge Univ. Press), 169
\bibitem[Howell et~al.(2013)]{How2013}{Howell}, D.~A., {Kasen}, D., {Lidman}, C., {et~al.} 2013, \apj, 779, 98
\bibitem[Inserra et~al.(2013)]{Inse2013}{Inserra}, C., {Smartt}, S.~J., {Jerkstrand}, A., {et~al.} 2013, \apj, 770, 128
\bibitem[Iwamoto et~al.(1998)]{Iwa1998}{Iwamoto}, K., {Mazzali}, P.~A., {Nomoto}, K., {et~al.} 1998, Natur, 395, 672
\bibitem[Janka et~al.(2007)]{Jan2007}{Janka}, H.-T., {Langanke}, K., {Marek}, A., {Martinez-Pinedo}, G., \& {M{\"u}ller}, B. 2007, Physics Reports, 442, 1-6, 38
\bibitem[Kasen \& Woosley(2009)]{Kas2009}{Kasen}, D., \& {Woosley}, S.~E. 2009, \apj, 703, 2205
\bibitem[Kasen \& Bildsten(2010)]{Kas2010}{Kasen}, D., \& {Bildsten}, L. 2010, \apj, 717, 245
\bibitem[Klein \& Chevalier(1978)]{Kle1978}{Klein}, R.~I., \& {Chevalier}, R.~A. 1978, ApJL, 223, L109
\bibitem[Kotera et~al.(2013)]{Kot2013}{Kotera}, K., {Phinney}, E.~S., \& {Olinto}, A.~V. 2013, \mnras, 432, 3228
\bibitem[Lundqvist et~al.(2001)]{Lun2001}{Lundqvist}, P., {Kozma}, C., {Sollerman}, J., \& {Fransson}, C. 2001, \aap, 374, 629
\bibitem[Lyman et~al.(2014)]{Lym2014}{Lyman}, J.~D., {Bersier}, D., {James}, P.~A., {Mazzali}, P.~A., {Eldridge}, J., {Fraser}, M., \& {Pian}, E. 2014, arXiv:1406.3667
\bibitem[MacFadyen \& Woosley(1999)]{Mac1999}{MacFadyen}, A.~I., \& {Woosley}, S.~E. 1999, \apj, 524, 262
\bibitem[Maeda et~al.(2003)]{Mae2003}{Maeda}, K., {Mazzali}, P.~A., {Deng}, J., {Nomoto}, K., {Yoshii}, Y., {Tomita}, H., \& {Kobayashi}, Y. 2003, \apj, 593, 931
\bibitem[Maeda et~al.(2007)]{Mae2007}{Maeda}, K., {Tanaka}, M., {Nomoto}, K., {Tominaga}, N., {Kawabata}, K., {Mazzali}, P.~A., {Umeda}, H., {Suzuki}, T., \& {Hattori}, T. 2007, \apj, 666, 1069
\bibitem[Mazzali et~al.(2000)]{Maz2000}{Mazzali}, P.~A., {Iwamoto}, K., \& {Nomoto}, K. 2000, \apj, 545, 407
\bibitem[Mazzali et~al.(2006)]{Maz2006}{Mazzali}, P.~A., {Deng}, J., {Pian}, E., {et~al.} 2006, \apj, 645, 1323
\bibitem[McCrum et~al.(2014)]{McC2014}{McCrum}, M., {Smartt}, S.~J., {Kotak}, R., {et~al.} 2014, \mnras, 437, 656
\bibitem[Metzger et~al.(2007)]{Met2007}{Metzger}, B.~D., {Thompson}, T.~A., \& {Quataert}, E. 2007, \apj, 659, 561
\bibitem[Metzger et~al.(2011)]{Met2011}{Metzger}, B.~D., {Giannios}, D., {Thompson}, T.~A., {Bucciantini}, N., \& {Quataert}, E. 2011, \mnras, 413, 2031
\bibitem[Moriya et~al.(2013)]{Mor2013}{Moriya}, T.~J., {Blinnikov}, S.~I., {Tominaga}, N., {Yoshida}, N., {Tanaka}, M., {Maeda}, K., \& {Nomoto}, K. 2013, \mnras, 428, 1020
\bibitem[Nakamura et~al.(2001)]{Nak2001}{Nakamura}, T., {Mazzali}, P.~A., {Nomoto}, K., \& {Iwamoto}, K. 2001, \apj, 550, 991
\bibitem[Nicholl et~al.(2013)]{Nich2013}{Nicholl}, M., {Smartt}, S.~J., {Jerkstrand}, A., {et~al.} 2013, Natur, 502, 346
\bibitem[Nicholl et~al.(2014)]{Nich2014}{Nicholl}, M., {Jerkstrand}, A., {Inserra}, C., {et~al.} 2014, \mnras, 444, 2096
\bibitem[Nugent et~al.(2011)]{Nug2011}{Nugent}, P.~E., {Sullivan}, M., {Cenko}, S.~B.,  {et~al.} 2011, Natur, 480, 344
\bibitem[Ofek et~al.(2013)]{Ofe2013}{Ofek}, E.~O., {Sullivan}, M., {Cenko}, S.~B., {et~al.} 2013, Natur, 494, 65
\bibitem[Ostriker \& Gunn(1971)]{Ost1971}{Ostriker}, J.~P., \& {Gunn}, J.~E. 1971, ApJL, 164, L95
\bibitem[Pastorello et~al.(2010)]{Pas2010}{Pastorello}, A., {Smartt}, S.~J., {Botticella}, M.~T., {et~al.} 2010, ApJL, 724, L16
\bibitem[Pejcha \& Thompson(2014)]{Pej2014}{Pejcha}, O., \& {Thompson}, T.~A. 2014, accepted to \apj, arXiv:1409.0540
\bibitem[Popov(1993)]{Pop1993}{Popov}, D.~V. 1993, \apj, 414, 712
\bibitem[Quimby et~al.(2011)]{Qui2011}{Quimby}, R.~M., {Kulkarni}, S.~R., {Kasliwal}, M.~M., {et~al.} 2011, Natur, 474, 487
and Gamma-Ray Bursts, ed. P. W.A. Roming, N. Kawai, \& E. Pian (Cambridge: Cambridge Univ. Press), 22
\bibitem[Sanders et~al.(2012)]{San2012}{Sanders}, N.~E., {Soderberg}, A.~M., {Valenti}, S., {et~al.} 2012, \apj, 756, 184
\bibitem[Soderberg et~al.(2005)]{Sod2005}{Soderberg}, A.~M.,  {Kulkarni}, S.~R., {Fox}, D. B., et al. 2005, \apj, 627, 877
\bibitem[Sollerman et~al.(2002)]{Sol2002}{Sollerman}, J., {Holland}, S.~T., {Challis}, P., {et~al.} 2002, \aap, 386, 944
\bibitem[Smith \& McCray(2007)]{Smi2007}{Smith}, N, \& {McCray}, R. 2007, ApJL, 671, L17
\bibitem[Stritzinger \& Leibundgut(2005)]{Str2005}{Stritzinger}, M., \& {Leibundgut}, B. 2005, \aap, 431, 423
\bibitem[Strolger et al.(2002)]{Str2002}{Strolger}, L.~G., {Smith}, R.~C., {Suntzeff}, N.~B., {et~al.} 2002, \aj, 124, 2905
\bibitem[Sutherland \& Wheeler(1984)]{Sut1984}{Sutherland}, P.~G., \& {Wheeler}, J.~C., 1984, \apj, 280, 282
\bibitem[Taddia et~al.(2015)]{Tad2015}{Taddia}, F., {Sollerman}, J., {Leloudas}, G., {Stritzinger}, M.~D., {Valenti}, S., {Galbany}, L., {Kessler}, R., {Schneider}, D.~P., \& {Wheeler}, J.~C.
 2015, \aap, 574, A60
\bibitem[Ugliano et~al.(2012)]{Ugl2012}{Ugliano}, M., {Janka}, H.-T., {Marek}, A., \& {Arcones}, A. 2012, \apj, 757, 69
\bibitem[Umeda \& Nomoto(2008)]{Ume2008}{Umeda}, H., \& {Nomoto}, K. 2008, \apj, 673, 1014
\bibitem[Usov(1992)]{Uso1992}{Usov}, V~V. 1992, Natur, 357, 472
\bibitem[Valenti et~al.(2008)]{Val2008}{Valenti}, S., {Benetti}, S., {Cappellaro}, E., {et~al.} 2008, \mnras, 383, 1485
\bibitem[Valenti et~al.(2011)]{Val2011}{Valenti}, S., {Fraser}, M., {Benetti}, S., {et~al.} 2011, \mnras, 416, 3138
\bibitem[Vreeswijk et~al.(2014)]{Vre2014}{Vreeswijk}, P.~M., {Savaglio}, S, {Gal-Yam}, A., {et~al.} 2014, \apj, 797, 24
\bibitem[Walker et~al.(2014)]{Wal2014}{Walker}, E.~S., {Mazzali}, P.~A., {Pian}, E., {et~al.} 2014, \mnras, 442, 2768
\bibitem[Wang et~al.(2015)]{Wang2015}{Wang}, S.~Q., {Wang}, L.~J., {Dai}, Z.~G., \& {Wu}, X.~F. 2015, \apj, 799, 107
\bibitem[Wheeler et~al.(2000)]{Whe2000}{Wheeler}, J.~C., {Yi}, I., {H{\"o}flich}, P., \& {Wang}, L. 2000, \apj, 537, 810
\bibitem[Wheeler et~al.(2014)]{Whe2014}{Wheeler}, J.~C., {Johnson}, V., \& {Clocchiatti}, A. 2014, arXiv:1411.5975
\bibitem[Woosley(1993)]{Woos1993}{Woosley}, S.~E. 1993, \apj, 405, 273
\bibitem[Woosley(2010)]{Woos2010}{Woosley}, S.~E. 2010, ApJL, 719, L204
\bibitem[Woosley et~al.(2002)]{Woo2002}{Woosley}, S.~E., {Heger}, A., \& {Weaver}, T.~A. 2002, Reviews of Modern Physics, 74, 4, 1015
\bibitem[Woosley \& Bloom(2006)]{Woo2006}{Woosley}, S.~E., \& {Bloom}, J.~S. 2006, \araa, 44, 507
\bibitem[Yuan et~al.(2010)]{Yuan2010}{Yuan}, F., {Quimby}, R.~M., {Wheeler}, J.~C., {et~al.} 2010, \apj, 715, 1338
\bibitem[Zhang \& M{\'e}sz{\'a}ros(2001)]{Zhang2001}{Zhang}, B., \& {M{\'e}sz{\'a}ros}, P. 2001, ApJL, 552, L35
\bibitem[Zhang et~al.(2012)]{Zhang2012}{Zhang}, T.~M., {Wang}, X.~F., {Wu}, C., {et~al.} 2012, \aj, 144, 131

\end{thebibliography}
\end{document}